\pdfoutput=1

\documentclass[twoside,twocolumn,9pt]{article}
\usepackage{extsizes}
\usepackage[super,sort&compress,comma]{natbib} 
\usepackage[version=3]{mhchem}
\usepackage[left=1.5cm, right=1.5cm, top=1.785cm, bottom=2.0cm]{geometry}
\usepackage{balance}
\usepackage{packages/widetext}
\usepackage{times,mathptmx}
\usepackage{sectsty}
\usepackage{graphicx} 
\usepackage{lastpage}
\usepackage[format=plain,justification=justified,singlelinecheck=false,font={stretch=1.125,small,sf},labelfont=bf,labelsep=space]{caption}
\usepackage{float}
\usepackage{fancyhdr}
\usepackage{fnpos}
\usepackage[english]{babel}
\usepackage{array}
\usepackage{droidsans}
\usepackage{charter}
\usepackage[T1]{fontenc}
\usepackage[usenames,dvipsnames]{xcolor}
\usepackage{setspace}
\usepackage[compact]{titlesec}
\usepackage{epstopdf}

\usepackage{graphicx}
\usepackage{dcolumn}
\usepackage{bm}
\usepackage{appendix}
\usepackage{changes} 
\DeclareMathAlphabet{\mathcal}{OMS}{cmsy}{m}{n}
\DeclareSymbolFont{newfont}{OML}{cmm}{m}{it}
\DeclareMathSymbol{\epsilon}{3}{newfont}{15}

\newcommand{\mr}[1]{\mathrm{#1}}
\newcommand{\ii}{\ensuremath{\mathrm{i}}}
\newcommand{\vect}[1]{\ensuremath{\mathbf{#1}}}
\newcommand{\vectsym}[1]{\ensuremath{\boldsymbol{#1}}}

\begin{document}


\makeFNbottom
\makeatletter
\renewcommand\LARGE{\@setfontsize\LARGE{15pt}{17}}
\renewcommand\Large{\@setfontsize\Large{12pt}{14}}
\renewcommand\large{\@setfontsize\large{10pt}{12}}
\renewcommand\footnotesize{\@setfontsize\footnotesize{7pt}{10}}
\makeatother

\renewcommand{\thefootnote}{\fnsymbol{footnote}}
\renewcommand\footnoterule{\vspace*{1pt}%
\hrule width 3.5in height 0.4pt \color{black}\vspace*{5pt}} 
\setcounter{secnumdepth}{5}

\makeatletter 
\renewcommand\@biblabel[1]{#1}            
\renewcommand\@makefntext[1]%
{\noindent\makebox[0pt][r]{\@thefnmark\,}#1}
\makeatother 
\renewcommand{\figurename}{\small{Fig.}~}
\sectionfont{\sffamily\Large}
\subsectionfont{\normalsize}
\subsubsectionfont{\bf}
\setstretch{1.125} 
\setlength{\skip\footins}{0.8cm}
\setlength{\footnotesep}{0.25cm}
\setlength{\jot}{10pt}
\titlespacing*{\section}{0pt}{4pt}{4pt}
\titlespacing*{\subsection}{0pt}{15pt}{1pt}

\fancyfoot{}
\fancyfoot[RO]{\footnotesize{\sffamily{ ~\textbar  \hspace{2pt}\thepage}}}
\fancyfoot[LE]{\footnotesize{\sffamily{\thepage~\textbar\hspace{3.45cm} }}}
\fancyhead{}
\renewcommand{\headrulewidth}{0pt} 
\renewcommand{\footrulewidth}{0pt}
\setlength{\arrayrulewidth}{1pt}
\setlength{\columnsep}{6.5mm}
\setlength\bibsep{1pt}

\makeatletter 
\newlength{\figrulesep} 
\setlength{\figrulesep}{0.5\textfloatsep} 

\newcommand{\topfigrule}{\vspace*{-1pt}%
\noindent{\rule[-\figrulesep]{\columnwidth}{1.5pt}} }

\newcommand{\botfigrule}{\vspace*{-2pt}%
\noindent{\rule[\figrulesep]{\columnwidth}{1.5pt}} }

\newcommand{\dblfigrule}{\vspace*{-1pt}%
\noindent{\rule[-\figrulesep]{\textwidth}{1.5pt}} }

\makeatother

\twocolumn[
  \begin{@twocolumnfalse}
\sffamily
\begin{tabular}{m{0cm} p{17.5cm} }

 & \noindent\LARGE{\textbf{Magneto-electrical orientation of lipid-coated graphitic micro-particles in solution$^\dag$}} \\
\vspace{0.3cm} & \vspace{0.3cm} \\

 & \noindent\large{Johnny Nguyen,\textit{$^{a}$} Sonia Contera,\textit{$^{b}$} and Isabel Llorente Garc\'{i}a\textit{$^{a}$}} \\

\vspace{0.2cm} & \vspace{0.2cm} \\

 & \noindent\normalsize{We demonstrate, for the first time, confinement of the orientation of micron-sized graphitic flakes to a well-defined plane. We orient and rotationally trap lipid-coated highly ordered pyrolytic graphite (HOPG) micro-flakes in aqueous solution using a combination of uniform magnetic and AC electric fields and exploiting the anisotropic diamagnetic and electrical properties of HOPG. Measuring the rotational Brownian fluctuations of individual oriented particles in rotational traps, we quantitatively determine the rotational trap stiffness, maximum applied torque and polarization anisotropy of the micro-flakes, as well as their dependency on the electric field frequency. Additionally, we quantify interactions of the micro-particles with adjacent glass surfaces with various surface treatments. We outline the various applications of this work, including torque sensing in biological systems.} \\

\end{tabular}

 \end{@twocolumnfalse} \vspace{0.6cm}

  ]

\renewcommand*\rmdefault{bch}\normalfont\upshape
\rmfamily
\section*{}
\vspace{-1cm}


\footnotetext{\textit{$^{a}$~Department of Physics and Astronomy, University College London, Gower St., London WC1E 6BT, UK.; E-mail: i.llorente-garcia@ucl.ac.uk}}
\footnotetext{\textit{$^{b}$~Clarendon Laboratory, University of Oxford, OX1 3PU, UK. }}

\footnotetext{\dag~Electronic Supplementary Information (ESI) available: videos of HOPG micro-flakes in solution in the presence of a vertical orienting magnetic field, rotating to align with a horizontal AC electric field. See DOI: 10.1039/b000000x/}



\section{\label{sec:Intro} Introduction}

Micro- and nano-particles of carbon such as graphene/graphite platelets and carbon nanotubes have unique electrical, magnetic, optical and mechanical properties that, combined with their biocompatible nature and their chemical and biological functionalization capability, make them very attractive for numerous applications. The controlled manipulation and orientation of such micro/nano-particles over macroscopic length-scales has interesting applications for batteries and energy storage devices \cite{magnAlignmentGraphiteBatteries}, for opto-electronic devices, including recently developed magnetically and electrically switched graphene-based liquid crystals \cite{magnAlignGrapheneOxideLC, grapheneGraphiteFlakeElectricAlign2013,grapheneOxideMonolayersElectrAlign2014,grapheneMonolayersLCelectrAlign2015}, and for the creation of novel artificial composite materials with tailored anisotropic properties such as conductive polymers and gels \cite{magnAlignGrapheneOxideGel}, material reinforcements \cite{superparamagnCoatingForAlignmtScience2012}, materials for thermal management solutions \cite{alignedGrapheneThermalMaterialVacuumFiltration2011}, hydrophobic coatings, infra-red absorbing coatings, etc. The controlled orientation of graphene-based micro/nano-particles can aid the synthesis of large-scale mono-crystalline graphene composites \cite{largeGrapheneSingleCrystals,GrapheneMacroAssemblies} and can lead to advances in template-mediated synthesis, for instance to induce the ordered deposition of organic molecules \cite{orderedOrganicMolecAssemblies}.

In this paper, we demonstrate, for the first time, confinement of the orientation of graphitic micro-flakes to a well-defined plane using a novel magneto-electrical approach for fully controlling micro-particle orientation. We orient and rotationally trap soluble, biocompatible, lipid-coated micro-particles of highly ordered pyrolytic graphite (HOPG) in aqueous solution. Our scheme takes advantage of the anisotropic diamagnetic and electrical properties of HOPG micro-flakes (graphene layer stacks) \cite{HOPGdefinitionBook} to orient them parallel to a plane defined by two perpendicular fields: a vertical static magnetic field ($\sim 240\,\mr{mT}$); and a horizontal, linearly polarized electric field ($\sim 2 \times 10^4 \,\mr{V/m}$) oscillating at frequencies above $10\,\mathrm{MHz}$. Our inexpensive set-up is made from one permanent magnet and two thin wire electrodes, involving no micro-fabrication. While we demonstrate our scheme on HOPG, the impact of our results extends to other forms of carbon-based micro/nano-particles.  

Current methods for large-scale vertical orientation of graphene-based particles such as, for instance, vacuum filtration \cite{alignedGrapheneThermalMaterialVacuumFiltration2011} and chemical vapour deposition \cite{veticalAlignGOwithFeOvapordeposition}, are involved and costly and do not provide full control of graphene plane orientation. As for magnetic and electrical orientation methods, only a handful of publications have reported the magnetic alignment of graphite/graphene micro/nano-particles in solution \cite{magnAlignmentGraphiteAstro,magnAlignmentGraphiteBatteries,magnAlignGrapheneOxideLC,magnAlignGrapheneOxideGel}, and the use of electric fields to orient these particles has turned out to be more challenging than originally expected. Only very recently, the use of alternating-current (AC) electric fields has been reported as a successful method for the orientation of graphite/graphene flakes \cite{grapheneGraphiteFlakeElectricAlign2013,grapheneOxideMonolayersElectrAlign2014,grapheneMonolayersLCelectrAlign2015} and carbon nanotubes \cite{CNTs_1998,CNTs_2012}. However, these methods also lack full control of particle orientation, since the use of a single orienting field (either magnetic or electric) still allows particles to freely rotate around the field direction. 

Our results present a number of key novelties with respect to previous work. By simultaneously applying magnetic and AC electric fields in different directions, we restrict micro-particle rotations and confine particle orientation to the plane defined by these fields. Furthermore, we track and analyse the rotational motion of individual micro-particles, as opposed to monitoring ensemble averages. We detect micro-particle orientation using robust image processing techniques which improve the detection accuracy in comparison to indirect methods reported in previous studies \cite{grapheneGraphiteFlakeElectricAlign2013,grapheneOxideMonolayersElectrAlign2014,grapheneMonolayersLCelectrAlign2015}. By tracking the rotation of the particles and analyzing their Brownian orientational fluctuations around equilibrium in the rotational traps, we quantitatively determine the rotational trap stiffness, maximum applied orientational torques and relevant polarization anisotropy factor for the micro-flakes, as well as how these depend on the frequency of the AC electric field. This is the first report of such measurements for lipid-coated HOPG micro-flakes. Additionally, we quantitatively characterize the interaction of the micro-flakes with nearby functionalized glass surfaces, reliably discriminating this interaction from the effect of the orienting fields in our analysis, and potentially opening a new route to measuring the strength of interfacial interactions. Crucially, rotational trapping (confinement to a plane) allows the powerful use of Brownian fluctuation analysis methods which enable precise quantitative measurements that cannot be found in previous studies. 

Our achievement of controlled orientation paves the way for advances in a wide range of applications. As well as the applications mentioned above, the manipulation of individual biocompatible graphitic micro-particles can open up new exciting possibilities for biological and chemical sensing \cite{grapheneChemicalSensors2012,grapheneBiosensors2011,grapheneChemicalBioSensorsReview2010}, and for fundamental biophysical and biochemical studies. Indeed, our ability to detect weak torques shows that our scheme could be applied to the precise sensing of biologically relevant torques. Our experiments contribute to the currently growing interest in the diamagnetic manipulation of micro/nano-particles \cite{magnManipulMicroNanoStructs2014}, particularly within a biological context, where most matter is diamagnetic. The recent demonstration of magnetically controlled nano-valves \cite{polymersomeMagnetoValves2014} is one excellent example of the enormous potential of diamagnetic manipulation.

\section{\label{sec:theory} Principles of magnetic and electrical orientation}

In our scheme, we first apply a vertical magnetic field in order to align the HOPG micro-flakes parallel to the field direction. Once aligned, the micro-flakes are still free to rotate around the magnetic field direction. We then apply an AC electric field perpendicular to the magnetic field to rotate the micro-flakes and constrain their orientation to the plane containing both the magnetic and electric fields (see Fig. \ref{fig:coordinatesBandE}). 

We define a fixed laboratory frame of reference with axes ($X$, $Y$, $Z$), where $Z$ is the vertical direction (Fig. \ref{fig:coordinatesBandE}). The static magnetic field $\vect{B}_0$ is along $Z$ and the electric field $\vect{E}_0$ oscillates along the horizontal $X$ direction. We also define a particle frame of reference with axes ($x$, $y$, $z$) fixed to the HOPG particle. The $x$-$y$ plane corresponds to the HOPG graphene planes and $z$ is normal to the graphene planes. 

\subsection{\label{sec:theory:magnetic} Magnetic alignment of lipid-coated HOPG}


The magnetic properties of graphite/graphene have rarely been exploited despite the well known diamagnetic nature of graphite \cite{TemperatDependenceGraphiteMagnSusceptib2,diamagnetismGraphite}. HOPG is highly magnetically anisotropic and one of the most strongly diamagnetic materials known, particularly along the direction perpendicular to the graphene planes (out-of-plane direction). Its out-of-plane ($\perp$) and in-plane ($\parallel$) volume magnetic susceptibilities are $\chi_{\perp}=-4.5 \times 10^{-4}$ and $\chi_{\parallel}=-8.5 \times 10^{-5}$, respectively, as measured by Simon \textit{et al.} \cite{GeimHOPGproperties}. For comparison, the magnetic susceptibility of water (also diamagnetic) is $\chi_\mr{water}\approx -9 \times 10^{-6}$. 

The magnetic manipulation of micro-particles in solution presents the advantages of being contactless, non-invasive, biocompatible, largely insensitive to the solvent's conductivity, ionic strength and pH, not accompanied by undesirable effects in solution (such as electrophoretic migration or electrochemical reactions in electro-manipulation), and cheap and simple thanks to the  availability of strong NdFeB permanent magnets \cite{magnManipulMicroNanoStructs2014}.\\

The effective induced magnetic moment that a particle with magnetic susceptibility tensor $\boldsymbol{\chi}_2$, immersed in a fluid with magnetic susceptibility tensor $\boldsymbol{\chi}_1$, experiences in the presence of a uniform, static magnetic field, $\vect{B}_0$, is \cite{TBJonesBook}:
\begin{equation}\label{eqn:meff}
	\vect{m}_{\mathrm{eff}} \approx \frac{V_2}{\mu_0}(\boldsymbol{\chi}_2-\boldsymbol{\chi}_1)\cdot\vect{B}_0 \,,
\end{equation}
where $\mu_0$ is the permeability of free space and $V_2$ is the volume of the particle. Eqn. (\ref{eqn:meff}) results from the fact that, for diamagnetic particles, the magnetic susceptibilities are very small ($<10^{-3}$ in absolute value) so that demagnetizing fields inside the particle are negligibly small. As a consequence of this, particle shape and geometry have a negligible effect on the effective magnetic moment induced on the particle. In the particle frame of reference, the volume magnetic susceptibility tensors for the anisotropic HOPG particles ($\boldsymbol{\chi}_2$) and for the isotropic fluid ($\boldsymbol{\chi}_1$) are expressed as: 
\begin{eqnarray}\label{eqn:susceptTensors}
\boldsymbol{\chi}_2 =
\begin{pmatrix}
\chi_{\parallel} &     0                    &    0                \\
        0        &     \chi_{\parallel}     &    0                \\
        0        &     0                    &  \chi_{\perp}   
\end{pmatrix}
,  & \hspace{0.2cm}
\boldsymbol{\chi}_1 =
\begin{pmatrix}
     \chi_1      &     0                    &    0                \\
        0        &     \chi_1               &    0                \\
        0        &     0                    &  \chi_1   
\end{pmatrix}
,
\end{eqnarray}
where $\chi_1 \approx \chi_\mr{water}$ is the isotropic magnetic susceptibility of the aqueous solution and $\chi_{\parallel}$ and $\chi_{\perp}$ are the in-plane and out-of-plane magnetic susceptibilities for HOPG, given in the first paragraph of this sub-section. 
\begin{figure}
\centering
  \includegraphics[width=1\columnwidth]{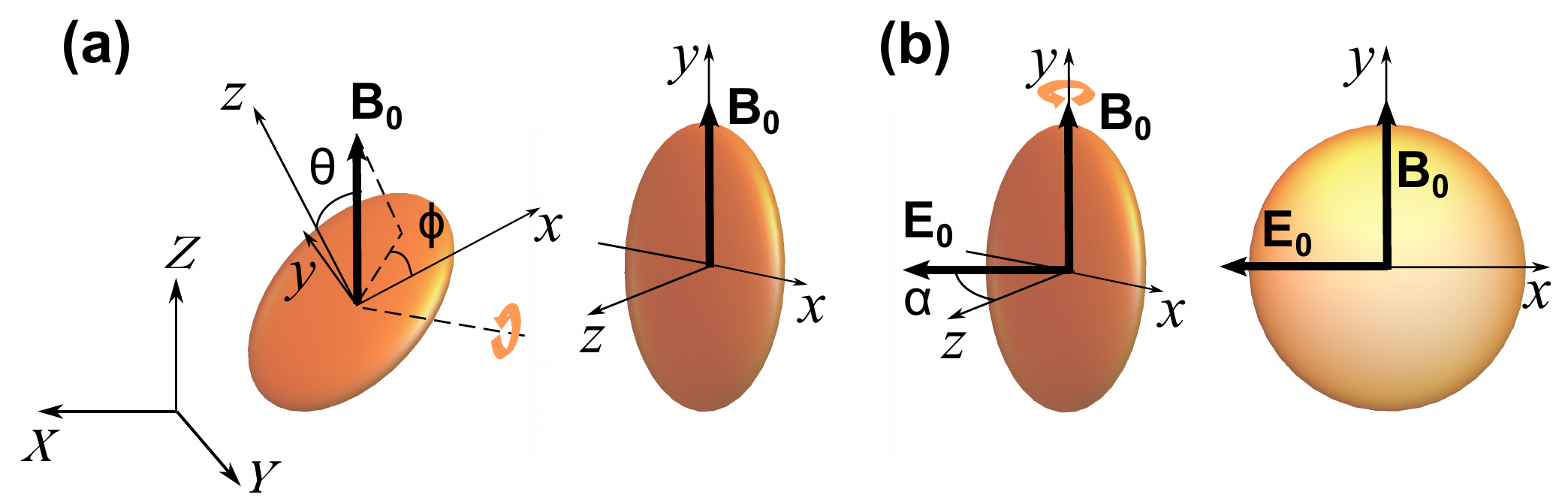}
  \caption{Schematic of magneto-electrical orientation of HOPG micro-flakes represented as oblate ellipsoids. (a) HOPG particle rotation upon application of a vertical magnetic field $\vect{B}_0$. (b) HOPG particle rotation upon additional application of a horizontal electric field $\vect{E}_0$.}
  \label{fig:coordinatesBandE}
\end{figure}

The direction of the magnetic field $\vect{B}_0$ in the particle frame is specified by the spherical polar angles ($\theta$,$\phi$), with $\theta$ being the angle with respect to $z$ and $\phi$ being the angle that the projection of $\vect{B}_0$ onto the $x$-$y$ plane makes with the $x$ axis [see Fig. \ref{fig:coordinatesBandE}(A)]. We can therefore write 
$\vect{B}_0 = B_0 (\sin\theta \cos\phi \,\hat{\vect{x}}+\sin\theta \sin\phi \,\hat{\vect{y}}+\cos\theta \,\hat{\vect{z}})$, where $B_0$ is the amplitude of the applied magnetic field. In the presence of this field, the particle experiences a magnetic torque $\vectsym{\mathcal{T}^\mr{m}}=\vect{m}_{\mathrm{eff}}\times\vect{B}_0$. The only non-zero components of this torque are $\mathcal{T}_x^\mr{m}$ and $\mathcal{T}_y^\mr{m}$, so that the particle will rotate around an in-plane axis until it reaches orientational equilibrium. The net magnetic torque around the in-plane ($\parallel$) axis is given by:
\begin{equation}\label{eqn:TxyMagnetic}
	\mathcal{T}_{\parallel}^\mr{m}=\frac{V_2 B_0^2}{2 \mu_0}(\chi_{\perp}-\chi_{\parallel})\sin(2\theta)\,.
\end{equation}
Note that this torque relies on the intrinsic magnetic anisotropy of the particles [$(\chi_{\perp}-\chi_{\parallel})\neq 0$)] and is independent of the magnetic susceptibility of the surrounding fluid. The torque depends on the orientation of the graphene planes with respect to the applied field and on particle volume, but is independent of particle shape.  This key aspect makes diamagnetic manipulation extremely versatile and powerful. Interestingly, the absolute value $\left| \chi_{\perp} - \chi_{\parallel} \right|$ for HOPG increases with decreasing temperature \cite{TemperatDependenceGraphiteMagnSusceptib1,TemperatDependenceGraphiteMagnSusceptib2,TemperatDependenceGraphiteMagnSusceptib3}.

The magnetic potential energy can be expressed as:
\begin{eqnarray}\label{eqn:Umagn}
	U^\mr{m} & = & -\int_{0}^{B_0} \vect{m}_{\mathrm{eff}}(B) \cdot \mr{d}\vect{B} \nonumber \\ 
	  & = & -\frac{V_2 B_0^2}{2 \mu_0} \left[ 
																						\left( \chi_{\parallel} - \chi_1       \right) +
																						\left( \chi_{\perp} - \chi_{\parallel} \right) \cos^2 \theta 
																		 \right] \, .
\end{eqnarray} 
The orientations of stable rotational equilibrium which satisfy $\partial U^\mr{m} / \partial \theta=0$ and $\partial^2 U^\mr{m} / \partial \theta^2 >0$ depend on the sign of the particle's magnetic anisotropy ($ \chi_{\perp} - \chi_{\parallel} $), so that for diamagnetic HOPG, with $( \chi_{\perp} - \chi_{\parallel} )<0$ and $\chi_{\perp} < \chi_{\parallel} < 0$, the equilibrium orientations correspond to $\theta = \pm 90 ^{\circ}$. Therefore, the HOPG flakes rotate until their graphene planes align parallel to the direction of the applied magnetic field, in order to minimize the magnetic interaction energy, as depicted in Fig. \ref{fig:coordinatesBandE}(A). 

The surrounding fluid has the effect of opposing the rotation of the particles through the contributions of rotational viscous drag and thermal rotational Brownian fluctuations. The magnetic field in our experiments ($B_0 \sim 240\,\mr{mT}$) is strong enough to overcome these contributions so that HOPG micro-particles can be quickly magnetically aligned and stably maintained into a vertical orientation. 

Our HOPG micro-flakes are coated with a phospholipid layer to facilitate dispersion in aqueous solution. Phospholipid molecules are also diamagnetically anisotropic and tend to align perpendicular to applied magnetic fields, as evidenced by the magnetic deformation of phospholipid bilayers and liposomes in strong magnetic fields ($>4\,\mr{T}$) \cite{diamagnLipids1,diamagnLipids2,diamagnLipids3,magnManipulMicroNanoStructs2014}. We neglect magnetic effects on our lipid layers given that we use modest magnetic fields up to $0.3\,\mr{T}$, that the magnetic anisotropy of similar phospholipid molecules \cite{diamagnLipids2,diamagnLipids3} is two orders of magnitude lower than the value of ($\chi_{\perp} - \chi_{\parallel}$) for HOPG, and that lipid layers on our particles are $\sim 100$ times thinner than the HOPG particle size. Note that, by contrast, the lipid layer plays a major role in the AC electrical orientation of the HOPG particles, as detailed in the following section. 

\subsection{\label{sec:theory:electric} Electrical alignment of lipid-coated HOPG}

The theory describing the electro-orientation of lipid-coated HOPG micro-particles is more complex than that for diamagnetic orientation because the electric de-polarization effects cannot be ignored, i.e., the electric field inside the particle cannot be approximated to be equal to the applied external electric field (this could be done in the diamagnetic case because $\chi \ll 1$). Lipid-coated HOPG micro-flakes can be modeled as oblate, layered, anisotropic ellipsoids and the theoretical frameworks by Jones \cite{TBJonesBook} and Asami \cite{AsamiEllipsoidShell} can be used to calculate the non-trivial full expression for the effective induced electric moment on the particles. The dependency on electric field frequency $f$ is introduced via complex permittivities $\epsilon$ that describe the relevant dielectric and conducting properties of the HOPG core, lipid layer and solution ($\epsilon = \varepsilon-\frac{\ii \sigma}{\omega \varepsilon_0}$, where $\varepsilon$ and $\sigma$ are the relevant static relative permittivity and conductivity, respectively, $\varepsilon_0$ is the permittivity of free space and $\omega = 2 \pi f$ is the angular frequency of the applied AC electric field). The reader can refer to Asami \textit{et al.} \cite{AsamiEllipsoidShell} for details of how Laplace's equation can be solved to find out the electrical potential outside a layered ellipsoid in order to derive the effective dipole moment of the submerged particle and its dependency on the frequency of the AC electric field (see also section \ref{sec:Experim:FreqDependence} for an intuitive explanation of how this frequency dependency arises). The effective electric dipole moment components for a layered ellipsoid can be expressed as:
\begin{equation}\label{eqn:peff}
	p_{\mr{eff},k} (t) = V_2 \varepsilon_1 \varepsilon_0  K_k E_{0k}(t)\, ,
\end{equation}
where $V_2$ is the particle's volume, $\varepsilon_\mr{1}$ is the relative static permittivity of the fluid, the sub-index $k$ indicates the $x$, $y$, $z$ directions in the particle frame of reference, $K_k$ are the complex effective polarization factors and $E_{0k}(t)$ are the components of the external AC electric field. Similarly to the magnetic case, the orientational electric torque can be obtained as $\vectsym{\mathcal{T}}^\mr{e}=\vect{p}_{\mr{eff}} \times \vect{E_0}$. 

In our experiments, $\vect{E_0}$ is applied once the micro-flakes have been pre-aligned with the vertical magnetic field. $\vect{E_0}$ is linearly polarised along the horizontal $X$ direction and makes an angle $\alpha$ to the $z$ axis [see Fig. \ref{fig:coordinatesBandE}(B)], so that $\vect{E}_0 = E_0 \left( \sin \alpha\,\hat{\vect{x}} + \cos \alpha\,\hat{\vect{z}} \right)$, where $E_0$ is the field amplitude. In the presence of this field, the particles feel a time-averaged electric torque around the in-plane $y$ direction, of the form:
\begin{equation}\label{eqn:timeAveragedTorqueComponents2}
	\mathcal{T}_{\parallel}^\mr{e} = 
	\frac{1}{4} V_2 \varepsilon_1 \varepsilon_0 E_0^2 \sin(2\alpha)  \mr{Re}\left[K_{\perp}-K_{\parallel}\right] \, ,
\end{equation}
where $K_{\parallel}\equiv K_{x,y}$ and $K_{\perp} \equiv K_z$ are the effective complex polarization factors for the in-plane and out-of-plane particle directions, respectively. These factors include the dependency on electric-field frequency, on particle shape and on dielectric and conductive properties of HOPG, lipids and solution. The polarization factors  have non-trivial forms, particularly for layered particles with an anisotropic core, as is the case here \cite{TBJonesBook}. Note that Eqn. (\ref{eqn:timeAveragedTorqueComponents2}) has a very similar form to the previous Eqn. (\ref{eqn:TxyMagnetic}) for the magnetic orientational torque. The maximum amplitude of the electrical orientational torque, which occurs at angles $\alpha=\pm 45^{\circ}$, is given by:
\begin{equation}\label{eqn:Tmax}
					\mathcal{T}_\mr{max}^\mr{e}= 
	\frac{V_2 \varepsilon_1 \varepsilon_0 E_0^2}{4} \mr{Re}\left[K_{\parallel}-K_{\perp}\right] \, .
\end{equation}
Hence, we can write $\mathcal{T}_{\parallel}^\mr{e} = -\mathcal{T}_\mr{max}^\mr{e} \sin(2\alpha)$. Analogously to Eqn. (\ref{eqn:Umagn}), the electric interaction potential energy is: 
\begin{equation}\label{eqn:Uelectr}
	U^\mr{e} = -\frac{V_2 \varepsilon_1 \varepsilon_0 E_0^2}{4} \mr{Re}\left[K_{\parallel} + \left(K_{\perp} - K_{\parallel}\right)\cos^2\alpha \right] \,.
\end{equation} 
For the parameters in our experiments, with $\mr{Re}[K_{\perp} - K_{\parallel}]<0$ and $\mr{Re}[K_{\parallel}]>\mr{Re}[K_{\perp}]>0$, stable rotational equilibrium takes place for orientation angles $\alpha = \pm 90^{\circ}$, i.e., particles rotate until their graphene planes align parallel to the applied electric field direction. Once aligned, the particles are rotationally trapped as long as the electric field is on. For small-angle deviations from orientational equilibrium, the rotational trap can be considered approximately harmonic, i.e., $\mathcal{T}_{\parallel}^\mr{e}(\alpha) = -\mathcal{T}_\mr{max}^\mr{e} \sin(2\alpha) \approx -k_\mr{e} \alpha$, where $k_\mr{e} = 2 \mathcal{T}_\mr{max}^\mr{e}$ is the electrical rotational trap stiffness. 

\section{\label{sec:Experim} Experimental demonstration}

\subsection{\label{sec:Experim:lipidcoating} Solubilized lipid-coated HOPG micro-particles}

HOPG micro-particles are very hydrophobic and strongly aggregate in aqueous solution. Amphiphilic lipid molecules, which form biological cell membranes, are good biocompatible candidates to coat HOPG micro-particles and disperse them in aqueous saline solution, as predicted by simulation for graphene \cite{grapheneLipids1}. Lipid monolayers have been shown to coat graphene oxide \cite{grapheneLipids2,grapheneLipids3} and graphene sheets \cite{grapheneLipids4}. The protocol that we have developed for functionalization can be found in Appendix \ref{App:LipidCoating}.

\subsection{\label{sec:Experim:setup} Experimental set-up}

\begin{figure}[h]
\centering
  \includegraphics[width=0.95\columnwidth]{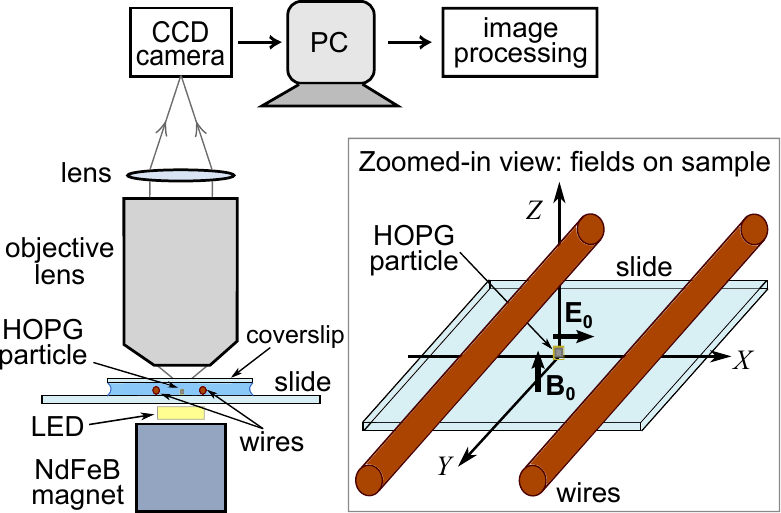}
  \caption{Schematic of experimental set-up. The magnet generates a vertical magnetic field along $Z$. The wires generate a horizontal electric field along $X$. HOPG micro-flakes are vertically oriented parallel to the $X$-$Z$ plane by the magnetic and electric fields as shown in the zoomed-in view and in more detail in Fig.\ref{fig:coordinatesBandE}. Micro-flake rotation is imaged from above the sample with a microscope onto a CCD camera (image plane parallel to $X$-$Y$ plane).}
  \label{fig:experimSetup}
\end{figure}

To generate a static, near-uniform magnetic field along the vertical $Z$ direction at the sample region, we use a NdFeB permanent magnet (grade N50, 25x25x10mm, $\sim 0.4\,\mr{T}$ at magnet surface, from Magnet Sales UK). The magnet is placed $\sim 6\,\mr{mm}$ below the sample (see Fig. \ref{fig:experimSetup}), resulting in a magnetic field strength $\sim 240\,\mr{mT}$ at the sample (measured with a gaussmeter). The magnetic field gradient along $Z$ is $\sim 0.03\,\mr{mT}/\mu\mr{m}$ so that variations in field strength and direction over the $\mu\mr{m}$ length scales relevant to our experiments are negligible. The horizontal time-varying electric field is generated by two parallel horizontal wires (lying along $Y$) glued onto the sample glass slide with nail varnish. These thin insulated copper wires ($50\,\mu\mr{m}$ diameter) are placed at a centre-to-centre distance $d \sim 150\,\mu\mr{m}$ (see Fig. \ref{fig:experimSetup}). An applied voltage of $\sim 4.6\,\mr{V_{pp}}$ (peak-to-peak) results in electric field magnitudes $E_0 \approx 2 \times 10^4\,\mr{V/m}$ at the sample region between the wires, with the field linearly polarized along the horizontal $X$ direction. We use AC electric field frequencies in the range $1-70\,\mr{MHz}$. A sample ($\sim 20 \,\mu\mr{l}$) of lipid-coated HOPG micro-particles in $20\,\mr{mM}$ NaCl aqueous solution is placed onto the glass slide with the two wires and sealed with a glass coverslip on top. The sample is imaged from above with a custom-made microscope (total magnification of 40) onto a CCD camera, at acquisition rates up to 150 frames per second. A small white-light LED illuminates the sample from below, as shown in Fig. \ref{fig:experimSetup}. 

We have carried out measurements of magneto-electrical orientation on 10 individual lipid-coated HOPG particles with a narrow spread of particle sizes and shapes and with average dimensions $\sim 2\,\mu\mr{m} \times 4\,\mu\mr{m} \times 7\,\mu\mr{m}$ ($0.5\,\mu\mr{m}$ standard deviations, see Appendix \ref{App:ImageProcessing}). The vertical $Z$ position of all micro-flakes imaged in experiments is approximately the same, with variations of order $0.5\,\mu\mr{m}$.

\subsection{\label{sec:Experim:Sequence} Measurement sequence}

\begin{figure}[h]
\centering
  \includegraphics[width=0.75\columnwidth]{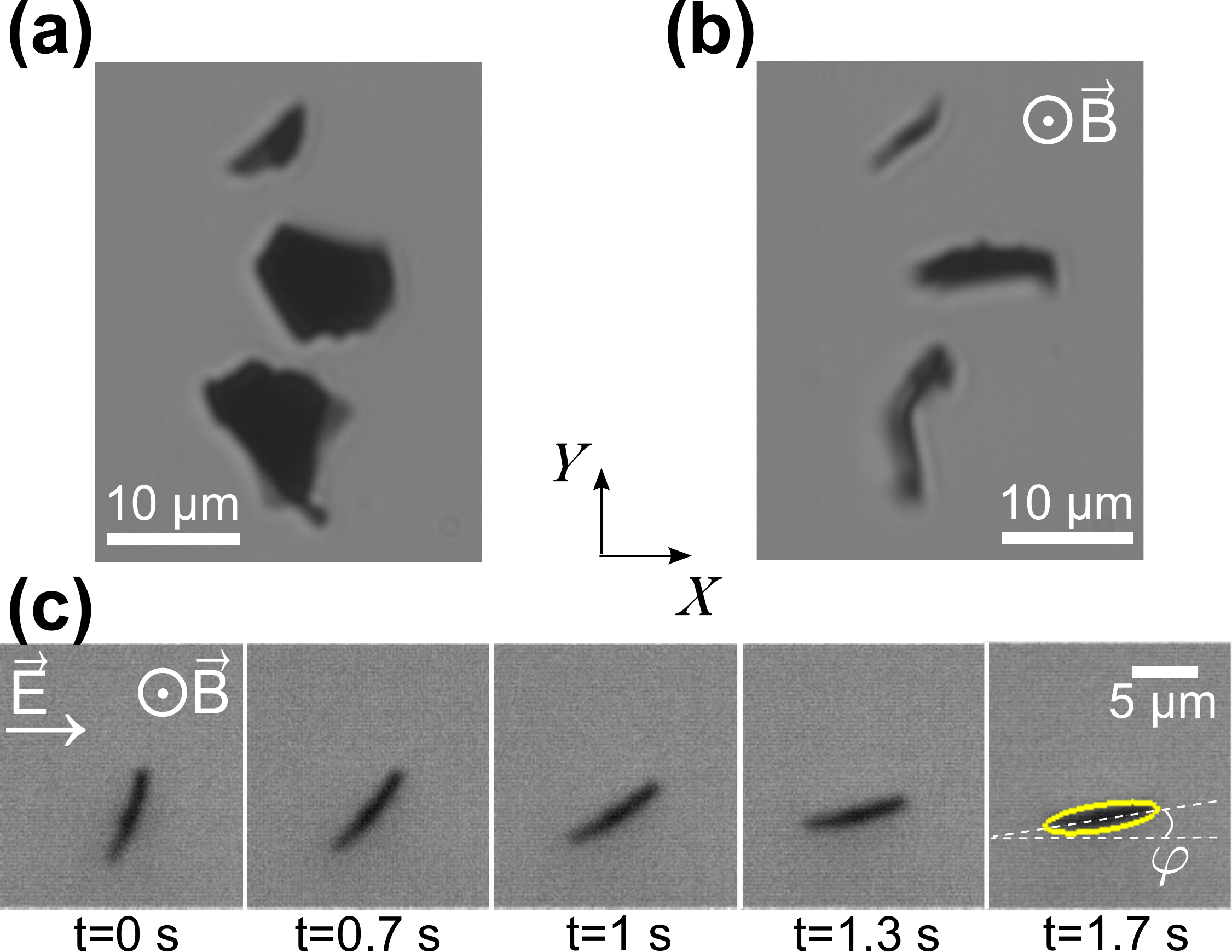}
  \caption{Microscope brightfield (transmission) images of lipid-coated HOPG micro-flakes: (a) with no applied fields; (b) vertically aligned in the presence of an applied vertical magnetic field $\sim 240\,\mr{mT}$; (c) sequence of rotation around $Z$ upon turning on a horizontal electric field oscillating at 30 MHz.}
  \label{fig:magneticOrientationImages}
\end{figure}

In the absence of applied fields, the HOPG micro-flakes in the sample solution tend to lie horizontally, with their graphene planes parallel to the slide surface, due to their oblate shape, as shown in Fig. \ref{fig:magneticOrientationImages}(A). When the magnet is placed under the sample, the vertical magnetic field aligns the micro-flakes onto vertical planes, as shown in Fig. \ref{fig:magneticOrientationImages}(B), following the principles explained in section \ref{sec:theory:magnetic}. At this point, the micro-flakes are still free to rotate around the vertical $Z$ axis. When the horizontal AC electric field is turned on, the micro-flakes rotate around $Z$ until their graphene planes align with the applied electric field direction as shown in the image sequence in Fig. \ref{fig:magneticOrientationImages}(C) and as described in section \ref{sec:theory:electric}. The micro-flake orientation is thereby confined to the $X$-$Z$ plane. The rotation of the micro-flakes when the electric field is applied is recorded with the microscope and camera (see Supplementary Videos). Individual isolated particles are imaged in order to avoid interaction effects. No heating effects are observed throughout the experiments.  

\subsection{\label{sec:Experim:ImageAnalysis} Image analysis: tracking rotational motion}

The use of the strong magnetic field to keep the micro-flakes vertically aligned allows us to restrict our analysis to a single rotational degree of freedom, $\varphi$, for rotations around $Z$,  where we define $\varphi$ as the angle between the HOPG graphene planes and the positive $X$ direction [see Fig. \ref{fig:magneticOrientationImages}(C)]. For our fixed electric field polarization along $X$, $\varphi=(90^{\circ}-\alpha)$ and $\mathcal{T}_{\parallel}^\mr{e} = -\mathcal{T}_\mr{max}^\mr{e} \sin(2\varphi)$ [see Eqn. (\ref{eqn:timeAveragedTorqueComponents2})], with orientational equilibrium corresponding to $\varphi = 0^{\circ}$.

The particle-rotation video sequences are analyzed using automated image processing algorithms in Matlab, in order to extract $\varphi$ and the particle dimensions as described in Appendix \ref{App:ImageProcessing}. This reproducible detection of the particle's orientation represents an improvement with respect to alternative indirect methods reported in the literature \cite{grapheneGraphiteFlakeElectricAlign2013,grapheneOxideMonolayersElectrAlign2014,grapheneMonolayersLCelectrAlign2015} based on light-transmission levels through graphitic micro-particle dispersions.

\subsection{\label{sec:Experim:FreqDependence} Frequency dependency - effect of the lipid coating}

The use of AC fields and insulated wires as electrodes is essential to avoid undesired electrochemical effects in ionic and biocompatible solutions. Furthermore, given that phospholipid membranes are insulating \cite{TBJonesBook}, MHz AC electric fields are required to observe the electro-orientation of the particles. This is because the lipid layer effectively insulates the HOPG at low and medium frequencies and only becomes electrically transparent at high frequencies. Maxwell-Wagner interfacial polarization effects arise due to the presence of several interfaces (solution-lipids-HOPG) with characteristic charge build-up and relaxation time constants which depend on the electrical properties at either side of the boundary \cite{TBJonesBook}. These time-varying interfacial free charges introduce a time dependency of the effective induced electrical dipole moment and give rise to a frequency dependency of the electric torque experienced by the lipid-coated HOPG particles, as explained in the theory section \ref{sec:theory:electric} and as evidenced by the measurements presented in section \ref{sec:results}.

\subsection{\label{sec:Experim:ImpedanceMatching} Actual electric field amplitude at the sample}

For a full quantitative discussion of the electric torque exerted on the micro-flakes, we need to know the exact magnitude of the electric field between the sample electrodes. At the high frequencies employed ($1-70\,\mr{MHz}$), electrical impedance-mismatch effects in the circuit connections can give rise to reflections at junction boundaries and cavity build-up effects. These lead to frequency-dependent variations of the effective electric field amplitude at the sample. The input voltage signal fed to our sample wires also presents some variation with frequency. These effects are measured and taken into consideration in order to correct our measured values of rotational trap stiffness and maximum electric torque as a function of frequency, as detailed in Appendix \ref{App:CorrectionFactor}.

\section{\label{sec:results} Results and discussion}

Below, we initially present data for the observed rotation of individual lipid-coated HOPG micro-flakes upon application of the AC electric field, including fits to the solution of the rotational equation of motion neglecting thermal fluctuations. We then use measurements of rotational Brownian fluctuations around the equilibrium orientation for aligned particles (rotationally trapped) to determine the rotational trap stiffness $k_\mr{e}$, maximum electric torque $\mathcal{T}_\mr{max}^\mr{e}$ and polarization anisotropy factor $\mr{Re}\left[K_{\parallel}-K_{\perp}\right]$ for the HOPG micro-flakes as a function of frequency. This analysis reveals the presence of weak interactions between the micro-particles and the nearby glass surface, which we also quantify to discriminate their effect from that of the electric torque.\\

All experiments are carried out in the presence of the vertical magnetic field $B_0 \approx 0.24 \,\mr{T}$. The maximum orienting magnetic torque [$V_2 B_0^2 (\chi_{\perp}-\chi_{\parallel})/(2 \mu_0)$ in Eqn. (\ref{eqn:TxyMagnetic})] applied to the particles is $\sim 2.4\times 10^{-16}\,\mr{N m}$, calculated for HOPG particles with average size $\sim 2\,\mu\mr{m} \times 4\,\mu\mr{m} \times 7\,\mu\mr{m}$ using the values of $\chi_{\parallel}$ and $\chi_{\perp}$ given in section \ref{sec:theory:magnetic}. This strong magnetic torque is two orders of magnitude larger than the applied electric torque (see measurements below) and is crucial to ensuring our micro-flakes stay vertically aligned thoughout the course of electro-orientation experiments.

\subsection{\label{sec:results:evolution} Rotation of micro-flakes}

\begin{figure}[h]
\centering
\includegraphics[width=0.9\columnwidth]{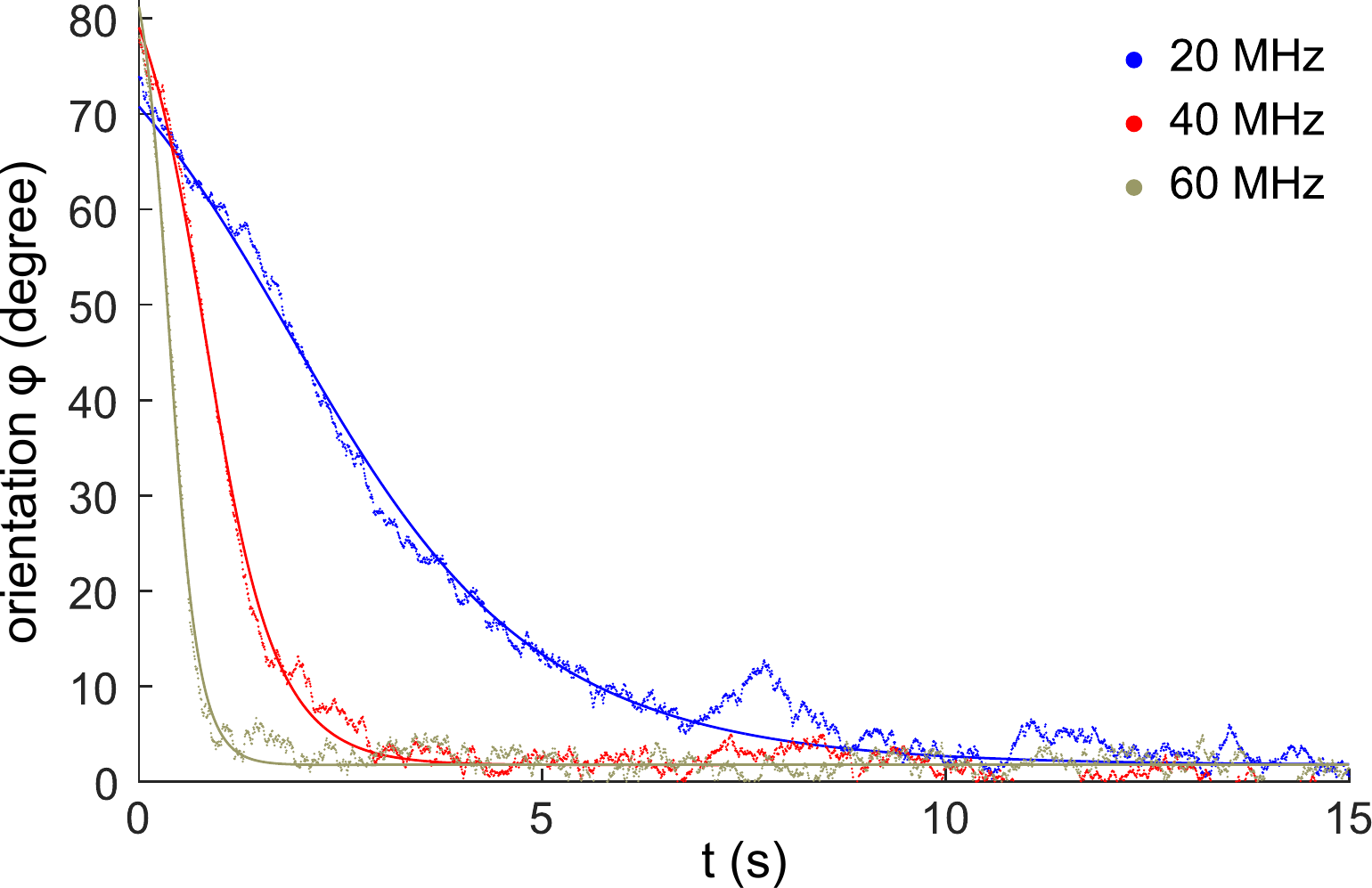}
	\caption{Measured rotation of a single HOPG micro-flake during electro-orientation. The electric field is turned on at $t=0$. The micro-flake rotates until it aligns with the electric field direction ($\varphi \approx 0^{\circ}$). Data shown for electric field frequencies: $20\,\mr{MHz}$, $40\,\mr{MHz}$ and $60\,\mr{MHz}$, with $E_0 \sim 2 \times 10^4 \,\mr{V/m}$. Solid lines: fits to Eqn. (\ref{eqn:solutionAngleVsTime}).}
	\label{fig:angleVsTime}
\end{figure}

Figure \ref{fig:angleVsTime} shows the measured evolution in time of the orientation angle $\varphi$ for a single micro-flake during electro-orientation (see also Supplementary Videos). Particle rotation is only observed for electric-field frequencies above $\sim 10\,\mr{MHz}$ (for $E_0 \sim 2 \times 10^4 \,\mr{V/m}$ and $V_2 \sim 30 \mu \mr{m}^3$ on average). Initially, $\varphi \approx 70^{\circ}-80^{\circ}$; the AC electric field is turned on at time zero and the micro-flake rotates until it reaches orientational equilibrium at $\varphi \approx 0^{\circ}$ within a few seconds. Figure \ref{fig:angleVsTime} shows that particles align faster with increasing electric field frequency, indicating stronger electric torques.

The equation of motion for the electro-orientation of the micro-particles in solution is given by the balance of torques:
\begin{equation}\label{eqn:FullRotEqnMotion}
					I_{\parallel}\ddot{\varphi}= -C_{\parallel}\dot{\varphi} + \mathcal{T}_{\parallel}^\mr{e} +
					\sqrt{2k_\mr{B} T C_{\parallel}}\, W(t) \, ,
\end{equation}
where $I_{\parallel}$ is the moment of inertia of the micro-flake around its in-plane axis, $C_{\parallel}$ is the corresponding rotational friction coefficient and $\mathcal{T}_{\parallel}^\mr{e}$ is the applied electric torque [$\mathcal{T}_{\parallel}^\mr{e} = -\mathcal{T}_\mr{max}^\mr{e} \sin(2\varphi)$, see Eqns. (\ref{eqn:timeAveragedTorqueComponents2}) and (\ref{eqn:Tmax})]. The last term in Eqn. (\ref{eqn:FullRotEqnMotion}) corresponds to the stochastic rotational Brownian fluctuations at temperature $T$, where $k_\mr{B}$ is Boltzmann's constant and $W(t)$ is a random variable with Gaussian distribution of zero mean and unity variance. 

The inertial term $I_{\parallel}\ddot{\varphi}$ in Eqn. (\ref{eqn:FullRotEqnMotion}) can be neglected in comparison to the viscous drag term $-C_{\parallel}\dot{\varphi}$. As a first approximation, we can neglect rotational Brownian fluctuations and solve $C_{\parallel}\dot{\varphi} = -\mathcal{T}_\mr{max}^\mr{e} \sin(2\varphi)$ to find an analytical expression for the evolution of the micro-flake orientation in time:
\begin{equation}\label{eqn:solutionAngleVsTime}
					\varphi(t) = \arctan \left[ \tan \left( \varphi_0 \right) \times \exp \left(- 2 \frac{\mathcal{T}_\mr{max}^\mr{e}}{C_{\parallel}} t \right)  \right] \, ,
\end{equation}
where $\varphi_0$ is the orientation angle at time $t=0$. Fits of our data to Eqn. (\ref{eqn:solutionAngleVsTime}) are presented in Figure \ref{fig:angleVsTime}, showing good agreement. The value of $\mathcal{T}_\mr{max}^\mr{e}$ could be extracted from these fits using calculated values for $C_{\parallel}$ (see Appendix \ref{App:OblateEllipsoidFormulas}). However, we use a more accurate method which does not rely on pre-calculated $C_{\parallel}$ values, as described in the following sub-section.

\subsection{\label{sec:results:Brownian} Orientational fluctuations for aligned micro-flakes}

By analysing orientational fluctuations of micro-flakes around equilibrium ($\varphi \approx 0^{\circ}$) in electrical rotational traps, we extract measurements of $\mathcal{T}_\mr{max}^\mr{e}$, $k_\mr{e}$ and $\mr{Re}\left[K_{\parallel}-K_{\perp}\right]$ at different electric field frequencies. Considering the effect of rotational Brownian fluctuations is essential in order to detect weak torques in environments dominated by thermal fluctuations. The Brownian motion of ellipsoidal particles was first theoretically described in 1934 by Perrin \cite{PerrinBrownianEllipsoid}, but was not experimentally verified until 2006 \cite{BrownianEllipsoidScience}.  

\begin{figure}[h]
\centering
	\includegraphics[width=1\columnwidth]{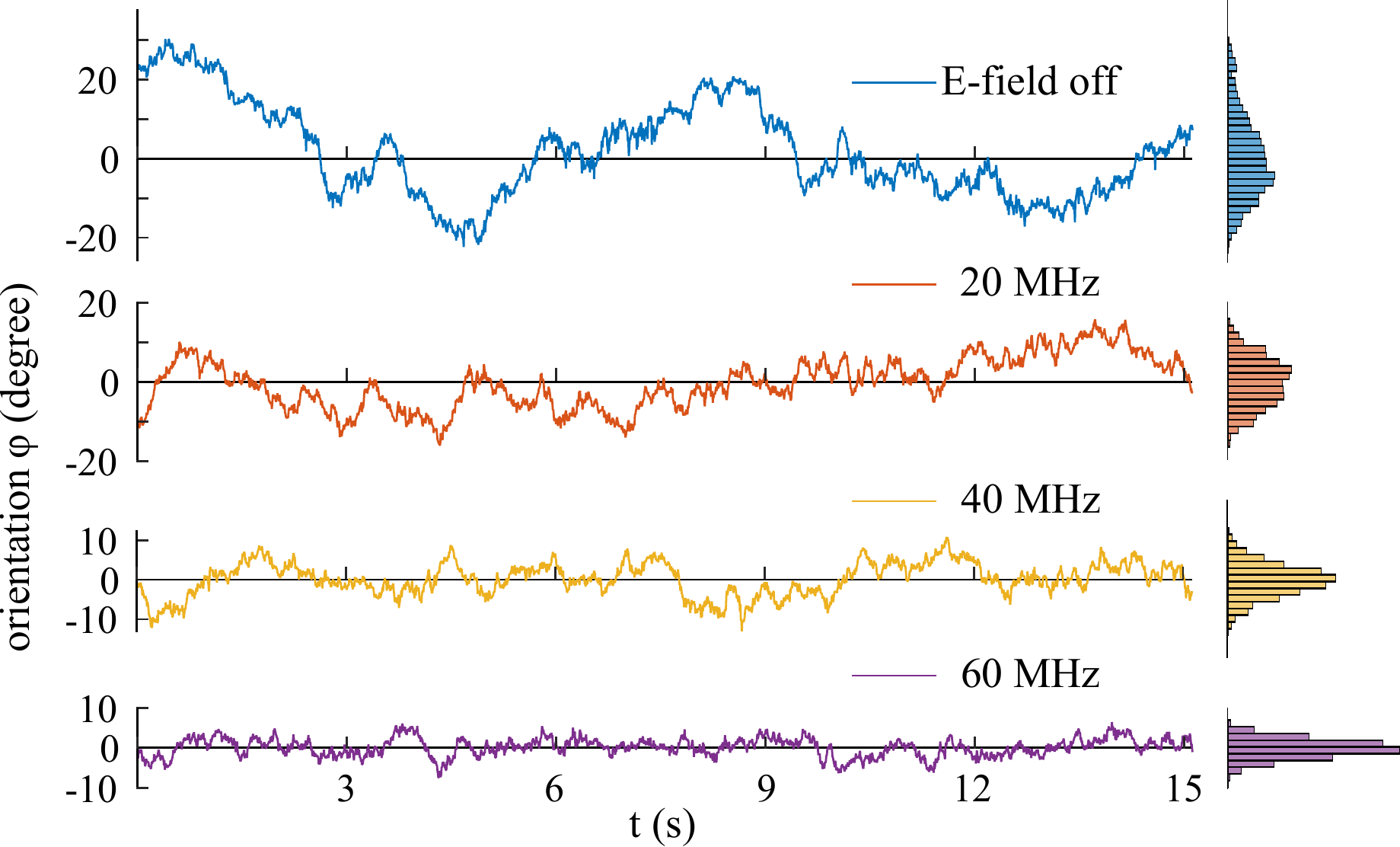}
	\caption{Fluctuations in HOPG micro-particle orientation $\varphi$ as a function of time in the absence and presence of the orienting electric field, for field frequencies $20\,\mr{MHz}$, $40\,\mr{MHz}$ and $60\,\mr{MHz}$. Histograms of $\varphi$ values shown on the right.}
	\label{fig:angularFlucts}
\end{figure}

The orientational fluctuations of the micro-flakes as they stay aligned with the electric field are monitored for 20-30 seconds. The electric field is then turned off and particles are recorded for a further 20-30 seconds to monitor their free rotational fluctuations in the absence of the electric field. Angular fluctuations derived from the measured video recordings are shown in Fig. \ref{fig:angularFlucts} for a single HOPG micro-flake at $\sim 23\,^{\circ}\mr{C}$. Data corresponding to electric field frequencies $20\,\mr{MHz}$, $40\,\mr{MHz}$ and $60\,\mr{MHz}$ are presented, together with data in the absence of the electric field. Fluctuations are clearly smaller in amplitude when the electric field is on, and decrease with increasing field frequency, as shown by the histograms on the right-hand side, evidencing rotational trapping and successful alignment. 

\begin{figure}[h]
\centering
\begin{minipage}{1\linewidth}
\centering
\includegraphics[width=0.95\textwidth]{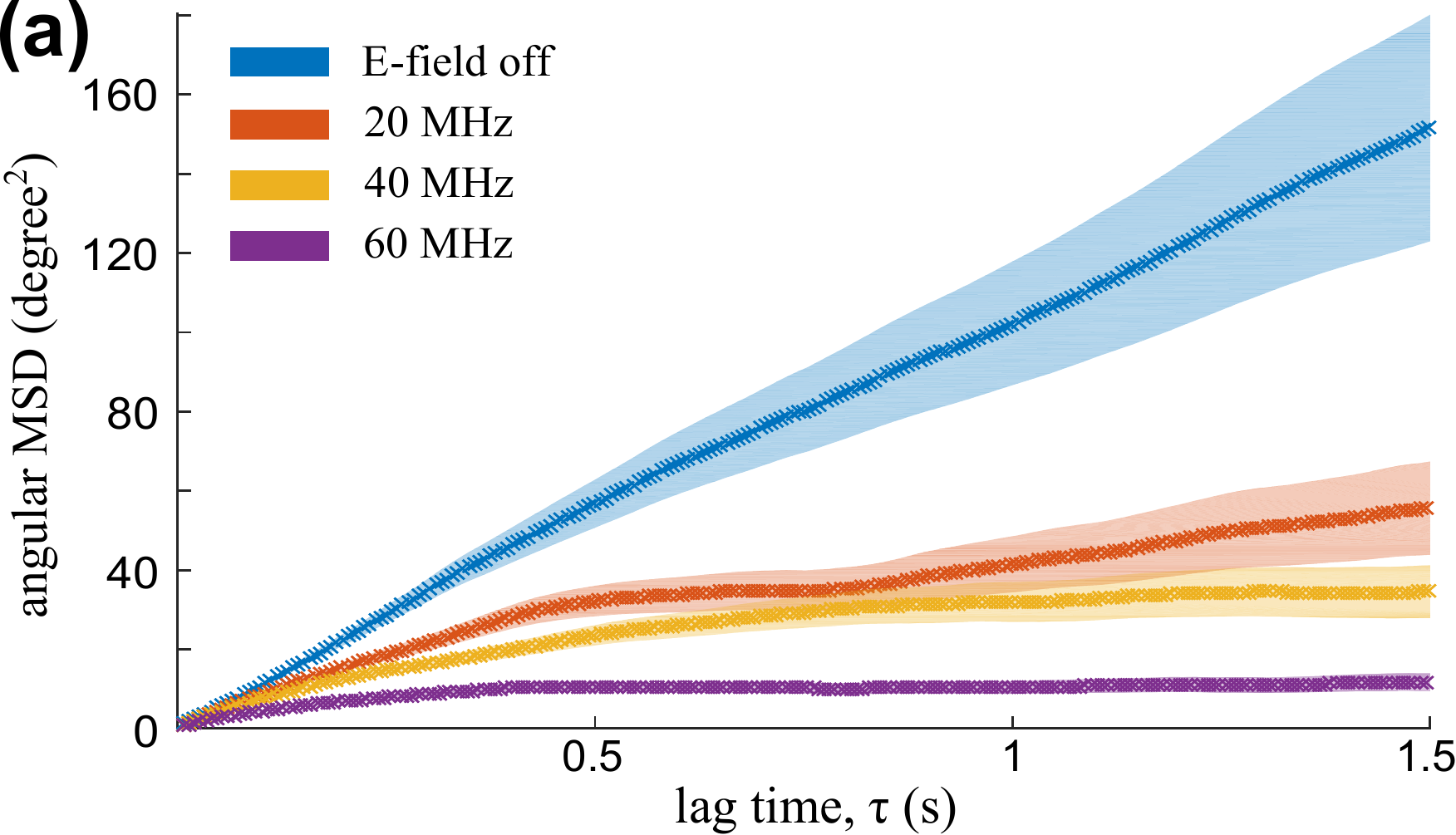}
\end{minipage}
\hspace{0.5cm}
\begin{minipage}{1\linewidth}
\centering
\includegraphics[width=0.95\textwidth]{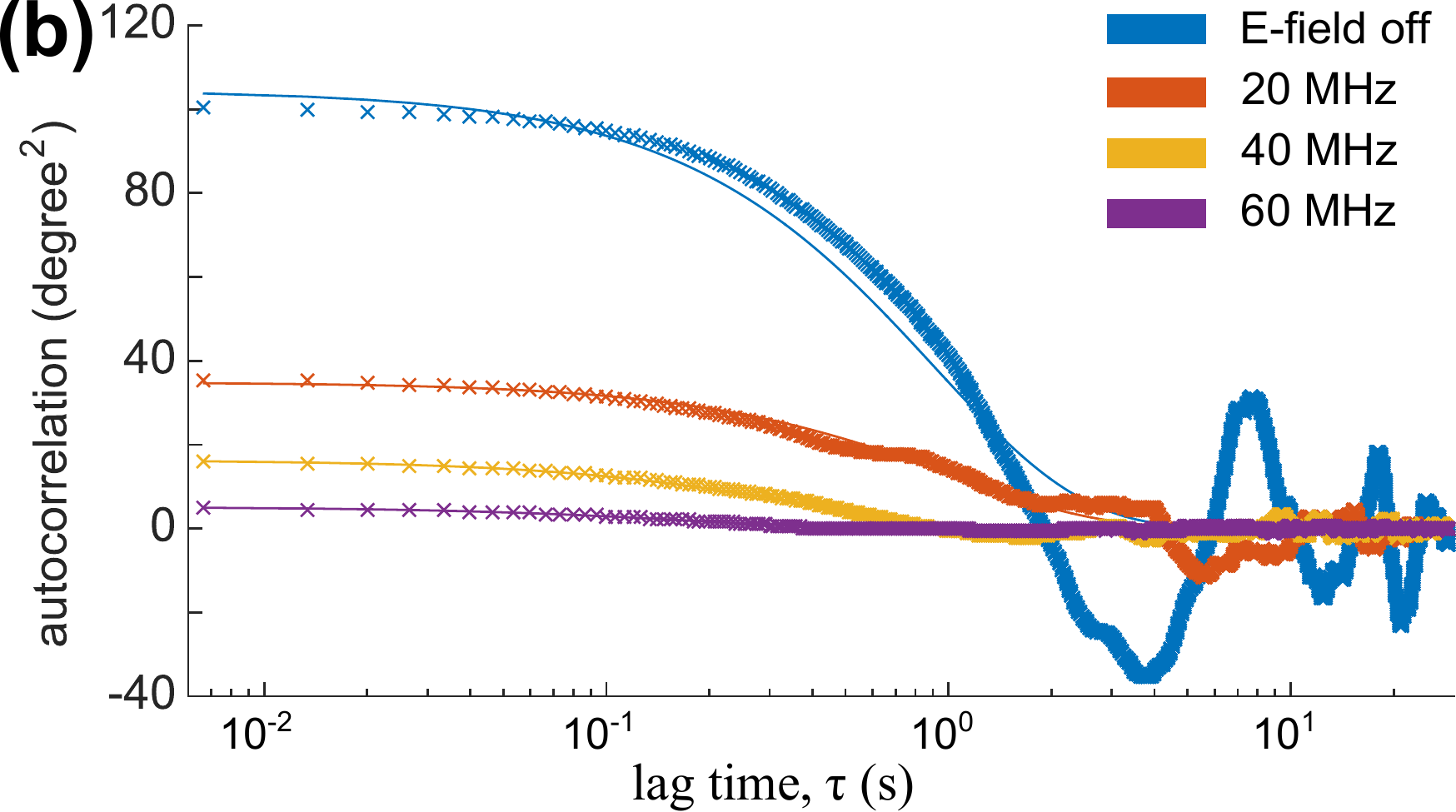}
\end{minipage}
\caption{(a) Angular MSD (data points and envelope for error bars) for the orientational fluctuation data shown in Fig. \ref{fig:angularFlucts}. (b) Auto-correlation [data points and fits to Eqn. (\ref{eqn:autocorr})] for the same fluctuation data.}
\label{fig:MSDandAutocorr}
\end{figure}

The mean square angular displacement (angular MSD) is defined as $\left\langle \left[ \Delta\varphi(\tau) \right]^2 \right\rangle$, where $\Delta \varphi (\tau) = \varphi(t_0+\tau)-\varphi(t_0)$ are the orientational fluctuations over a time interval $\tau$ and $t_0$ is the initial time. The angle brackets indicate averaging over all initial instants. For the case of free (untrapped) rotational diffusive behaviour, the angular MSD should depend linearly on $\tau$ as $\left\langle \left[ \Delta\varphi(\tau) \right]^2 \right\rangle = 2 D_{\varphi} \tau$, where $D_{\varphi} = k_\mr{B} T / C_{\parallel}$ is the corresponding rotational diffusion coefficient \cite{PerrinBrownianEllipsoid}. In the presence of rotational trapping, the angular MSD deviates from this linear behaviour. This is shown in Fig. \ref{fig:MSDandAutocorr}(A), where the measured angular MSD corresponding to the orientational fluctuation data in Fig. \ref{fig:angularFlucts} is presented. 

The non-linearity of the angular MSD due to rotational trapping is observed even when the electric field is off [Fig. \ref{fig:MSDandAutocorr}(A)]. This is due to the fact that particles fall under gravity and interact non-negligibly with the nearby glass surface in a way that opposes their rotation. Both the small effect of these interactions with the glass and the strong alignment in the presence of the AC electric field are described theoretically as rotational traps with trap stiffness $k_\mr{glass}$ and $k_\mr{e}$, respectively. For small-angle fluctuations, the electrical rotational trap is approximately harmonic so that $\mathcal{T}_{\parallel}^\mr{e}(\varphi) = -\mathcal{T}_\mr{max}^\mr{e} \sin(2\varphi) \approx -k_\mr{e} \varphi$, with $k_\mr{e} = 2 \mathcal{T}_\mr{max}^\mr{e}$. A similar assumption is made for the trap due to interactions with the glass. Hence, the full equation of rotational motion becomes:
\begin{equation}\label{eqn:FullRotEqnMotion2}
					-C_{\parallel}\dot{\varphi} - k_\mr{total} \varphi +
					\sqrt{2k_\mr{B} T C_{\parallel}}\, W(t) = 0\, ,
\end{equation}
where $k_\mr{total}$ is equal to either $(k_\mr{e}+k_\mr{glass})$ when the electric field is on, or $k_\mr{glass}$ when it is off. The well-known theory of Brownian motion in a harmonic trap in solution \cite{TheoryBrownianMotionII,OpticalTweezersPhilJones,BrownianMSDintrap} can be applied so that the angular MSD is given by:
\begin{equation}\label{eqn:angularMSDtrap}
					\left\langle \left[ \varphi(t_0+\tau)-\varphi(t_0) \right]^2 \right\rangle = 
					\frac{2 k_\mr{B}T}{k_\mr{total}} \left[ 1- \exp {\left( -\frac{k_\mr{total}}{C_{\parallel}} 					\tau \right)} \right] \, ,
\end{equation}
and the auto-correlation of the fluctuations around the equilibrium orientation in the rotational trap is:  
\begin{equation}\label{eqn:autocorr}
					R(\tau) = \left\langle \varphi(t_0) \varphi(t_0+\tau) \right\rangle = \frac{k_\mr{B}T}{k_\mr{total}} \exp{ \left(-\frac{k_\mr{total}}{C_{\parallel}} \tau \right)} \, .
\end{equation}

The angular MSD data in Fig. \ref{fig:MSDandAutocorr}(A) shows how the measured asymptotic values at long lag times [pre-factor $2 k_\mr{B}T / k_\mr{total}$in Eqn. (\ref{eqn:angularMSDtrap})] decrease with increasing electric field frequency, indicating increased rotational trap stiffness and electric torques at higher frequencies. The corresponding measured auto-correlation $R(\tau)$ is shown in Fig. \ref{fig:MSDandAutocorr}(B). The same trend of increasing trap stiffness with frequency is evidenced by the faster exponential decay and lower $R(\tau=0)$ values observed at higher frequencies, as given by Eqn. (\ref{eqn:autocorr}).\\

\subsubsection{Frequency dependency of maximum electric torque, rotational trap stiffness and polarisation anisotropy}

In order to extract quantitative values of $k_\mr{total}$, we fit the auto-correlation of the measured orientational fluctuations as a function of time lag to the exponential function in Eqn. (\ref{eqn:autocorr}). This choice is motivated by the fact that the equivalent method for translational fluctuations is considered one of the most reliable ones out of several available methods for calibrating the trap stiffness in optical tweezers \cite{OpticalTweezersPhilJones}. From the fit, $k_\mr{total}$ can be obtained from the auto-correlation at zero lag time $R(\tau=0)$, using $T = 23\,^{\circ}\mr{C}$. For each frequency, we extract $k_\mr{total,on}=(k_\mr{e}+k_\mr{glass})$ from the fluctuations recorded with the electric field on, and $k_\mr{total,off}=k_\mr{glass}$ from the fluctuations recorded with the electric field off to isolate interactions with the glass surface. The electrical trap stiffness, $k_\mr{e}$, is obtained from the subtraction $k_\mr{total,on}- k_\mr{total,off}$. The value of the maximum electric torque is obtained as $\mathcal{T}_\mr{max}^\mr{e} = k_\mr{e}/2$.

For a given micro-flake, approximately five measurements are taken at a given frequency, which are averaged to obtain mean and standard deviation values at that frequency. The standard errors of the mean are used as error bars. The process is repeated for different frequencies in order to investigate the frequency dependency of the electric torque, which follows from Eqn. (\ref{eqn:Tmax}). Results are then corrected for the voltage variations with frequency that originate from the RF source and impedance mismatch effects explained in section \ref{sec:Experim:ImpedanceMatching} and Appendix \ref{App:CorrectionFactor}.

\begin{figure}[h]
\centering
  \includegraphics[width=0.9\columnwidth]{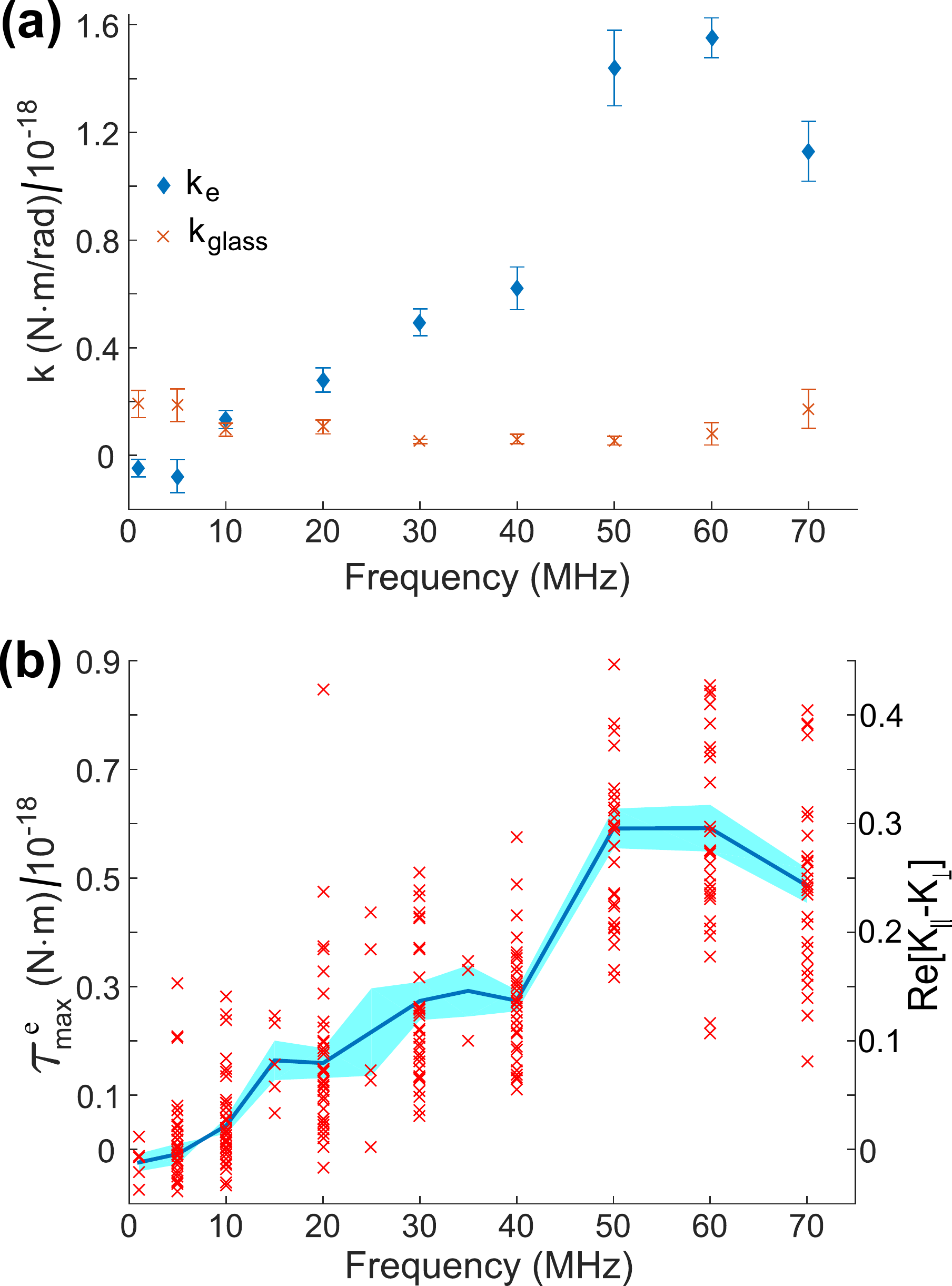}\\
  \caption{(a) Measured rotational trap stiffness versus frequency for an individual lipid-coated HOPG micro-flake: $k_\mr{e}$ for the electrical rotational trap and $k_\mr{glass}$ for the trap due to glass-particle interactions. Data points: mean value from 5-10 measurements per frequency. Error bars: standard error of the mean (SE). (b) Data for the maximum electric torque $\mathcal{T}_\mr{max}^\mr{e} = k_\mr{e}/2$ measured for 10 different micro-flakes. Data points from all particles shown together with overall mean value (solid line) and SE (light blue envelope) for each frequency. The axis on the right hand side shows the polarisation anisotropy factor $\mr{Re}\left[K_{\parallel}-K_{\perp}\right]$.}
  \label{fig:KversusFreq}
\end{figure}

Figure \ref{fig:KversusFreq}(A) shows values of $k_\mr{e}$ and $k_\mr{glass}$ versus frequency for the electro-orientation of an individual lipid-coated HOPG micro-flake in solution near a glass surface passivated with polyethylene glycol (PEG)-silane. PEGylation, i.e., the coating of a surface with largely non-interacting PEG polymer chain brushes, is a well known technique for surface passivation \cite{PEGcoating}. The resulting $k_\mr{glass}$ is low and approximately constant in time (independent of frequency, as expected), which enables the reliable characterisation of interactions between micro-particle and glass. The electric rotational trap stiffness $k_\mr{e}$, and therefore $\mathcal{T}_\mr{max}^\mr{e}$, decrease with decreasing frequency due to the electrically insulating effect of the lipid coating at low to medium frequencies. Data for the maximum electrical torque $\mathcal{T}_\mr{max}^\mr{e}$ measured for 10 different micro-flakes are presented in Fig. \ref{fig:KversusFreq}(B), including the overall mean and standard error for all measurements. Given the narrow spread of particle sizes and shapes chosen for our experiments, the trend with frequency remains clearly visible. From the 10 particles measured, 3 were on PEG-silane passivated glass slides, 4 on plasma cleaned slides and 3 on untreated glass.

In Fig. \ref{fig:KversusFreq}(B), the measured average $\mathcal{T}_\mr{max}^\mr{e}$ is in the range $0-0.6 \times 10^{-18}\,\mr{N m}$, depending on the applied electric field frequency, corresponding to an average torsional stiffness ($k_\mr{e}$) range of $0-1.2 \times 10^{-18}\,\mr{N m/rad}$. In the theoretical expression for $\mathcal{T}_\mr{max}^\mr{e}$ in Eqn. (\ref{eqn:Tmax}), we have a frequency-independent pre-factor, $V_2 \varepsilon_1 \varepsilon_0 E_0^2/4$, multiplying the frequency-dependent polarization anisotropy factor $\mr{Re}\left[K_{\parallel}-K_{\perp}\right]$. The pre-factor evaluates to $\sim 2\times 10^{-18}\,\mr{N m}$, calculated using $V_2=4 \pi a b c/3$ for our particles modeled as ellipsoids with semi-axis lengths $a=1\,\mu\mr{m}$, $b=2\,\mu\mr{m}$ and $c=3.5\,\mu\mr{m}$ on average, $\varepsilon_0=8.85\times 10^{-12}\,\mr{F/m}$, $\varepsilon_1\approx 80$ for $20\,\mr{mM}$ NaCl at $23\,^{\circ}\mr{C}$ \cite{NaClsolutionProps} and $E_0=2\times 10^4\,\mr{V/m}$. Using the $\mathcal{T}_\mr{max}^\mr{e}$ measurements in Fig. \ref{fig:KversusFreq}(B) and this calculated pre-factor we obtain a polarization anisotropy factor in the range $\sim 0-0.3$ for our lipid-coated HOPG micro-flakes, for frequencies in the range 10-70 MHz. This is indicated by the right-hand-side axis in Fig. \ref{fig:KversusFreq}(B) and is a useful measurement for a number of HOPG electro-manipulation experiments \cite{TBJonesBook}. 

The reproducible frequency dependence of the orientational torque on the micro-flakes is a useful feature which enables frequency-control of micro-particle orientation for numerous applications. The turnover frequency at which electric torque values begin to increase can be altered, for instance, by modifying the thickness of the insulating lipid layer on the micro-flakes, the conductivity of the solution or the micro-flake aspect ratio, opening interesting control possibilities.\\

\subsubsection{Interactions with the glass surface}

Additionally, we have used our measurements to characterize the interactions of individual lipid-coated HOPG micro-flakes with glass slides with various surface treatments, given that micro-flakes fall under gravity coming into contact with the glass slide. The quantitative characterization of surface sticking effects and glass-lipid interactions is of particular interest to applications in biology. 
\begin{figure}[h]
\centering
  \includegraphics[width=0.8\columnwidth]{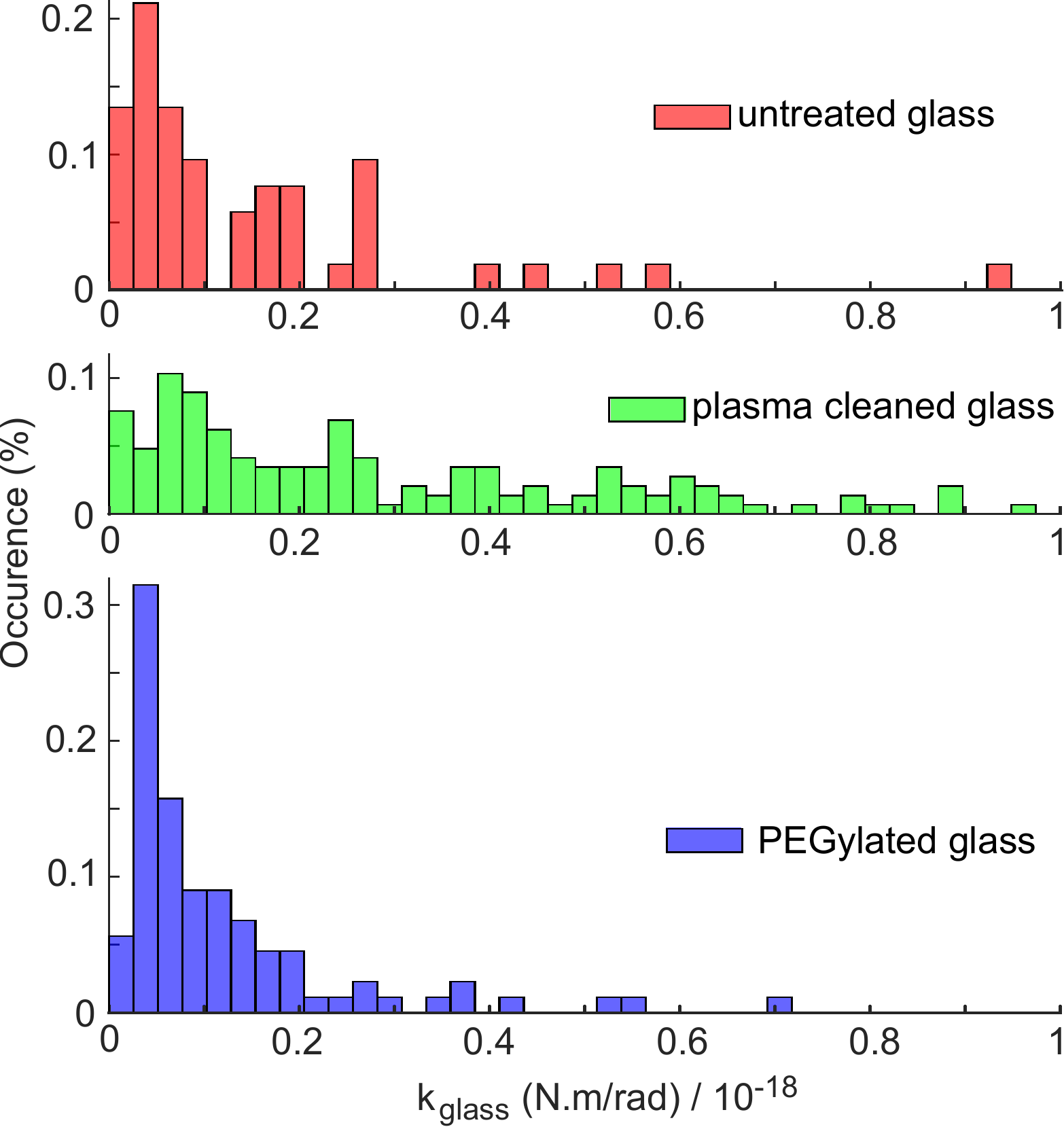}\\
  \caption{Characterization of interactions of individual lipid-coated HOPG micro-flakes with glass surfaces with various treatments. Histograms of trap stiffness $k_\mr{glass}$ due to glass-particle interactions for 10 different particles on untreated glass slides (48 measurements), plasma cleaned glass slides (145 measurements) and PEG-silane passivated glass slides (89 measurements).}
  \label{fig:kglassHist}
\end{figure}

$k_\mr{glass}$ is determined from the above mentioned measurements of rotational Brownian fluctuations acquired in the absence of the electric field as explained in the previous paragraphs in section \ref{sec:results:Brownian}. Results for $k_\mr{glass}$ from 282 measurements for 10 different particles on untreated glass, plasma-cleaned glass and PEG-silane passivated glass are compared in the histograms shown in Fig. \ref{fig:kglassHist}. The histograms present clear differences, with PEG-silane passivation leading to the lowest $k_\mr{glass}$ mean and spread values, i.e., to the weakest rotational trapping due to surface sticking. Untreated glass shows a wider distribution of $k_\mr{glass}$ values than PEG-silane treated glass. Plasma-cleaned slides show an even larger $k_\mr{glass}$ spread, as expected by the generation of charged groups on the glass surface during the plasma cleaning process. While measurements have been carried out using 10 different particles with a narrow spread of particle sizes and shapes (see Appendix B), differences in particle geometry and the various possible values of glass-particle contact area throughout the measurements contribute to the spread of the measured $k_\mr{glass}$ values.

\section{\label{sec:Conclusions} Conclusions and outlook}

We have presented a detailed study of the magneto-electrical orientation and rotational trapping of lipid-coated HOPG micro-particles in aqueous solution, including a solid theoretical framework and quantitative experimental results. Measurements of the maximum magnetic and electric orientational torques, rotational trap stiffness and polarisation anisotropy of the micro-particles have been presented and their dependency on the frequency of the applied electric field has been investigated. These measurements are the first reported for lipid-coated HOPG micro-flakes, with the observed frequency dependency opening the door to the implementation of frequency controlled electro-orientation using this technique. The interactions of the lipid-coated particles with glass surfaces with different surface treatments have been characterized by analyzing their weak rotational trapping effect on the micro-flakes, with this effect being reliably discriminated from the stronger rotational trapping generated by the electric field.  

Our results demonstrate, for the first time, confinement to a well-defined plane of the orientation of graphitic micro-flakes in solution, with this new method exploiting the electrical and diamagnetic anisotropy of the particles via the application of simultaneous perpendicular magnetic and electric fields. The combination of magnetic and electric fields in different configurations can open up new ways of manipulating graphitic micro/nano-particles for their orientation, confinement and transport \cite{TBJonesBook,DEPinBioReview}, with the use of time-varying fields enabling frequency-control of the magneto-electrical manipulation. The principles of our method extend to other carbon-based micro/nano-particles, such as graphene platelets or carbon-nanotubes, and our scheme has great potential for being scaled down via micro-fabrication. Alternative schemes could be devised to orient and confine graphitic micro-flakes to a plane making use of rotating electric/magnetic fields, or of fast time-switching between fields in two perpendicular directions, as demonstrated e.g. for the orientation of diamagnetic, anisotropic polymer and cellulose fibers \cite{orientPolymerFibers,orientCelluloseFibers}. 

Applications in biochemistry and biological and medical physics are particularly relevant to our experiments. Our lipid-coated HOPG micro-particles are biocompatible and our results in NaCl aqueous solution easily extend to biocompatible solutions. Our particles can be functionalized and specifically bound to biomolecules such as antibodies, protein complexes and nucleic acids, and the lipids can be fluorescently labeled or conjugated to polymers, for instance to change the electrical properties of the coating \cite{electricallyAddressableVesicles}. HOPG micro-particles of regular sizes could be generated using current micro-fabrication techniques, such as ion-beam milling. Our particles hence constitute a promising tool with the potential to function as carriers, labels or specific targets for biological and chemical sensing applications \cite{newMaterialsElectrochemSensingBeads2001,grapheneChemicalSensors2012,grapheneBiosensors2011,grapheneChemicalBioSensorsReview2010}. Cells could be attached onto HOPG micro-plates which could be manipulated for versatile cell-to-cell interaction experiments, such as those involving immune response, virus transfer, neuron activity, etc. As another example, given their capability of absorbing infra-red light, graphitic micro-particles could also be good candidates for photo-thermal cell therapy (e.g. for cancer treatment) or for temperature-jump studies \textit{in vitro}/\textit{in vivo}.    

Furthermore, our HOPG micro-particles in calibrated rotational traps of known torsional trap stiffness could be used for sensing biologically relevant torques. These are typically in the range $0.01-1 \times 10^{-18}\,\mr{N m}$ \cite{torqueSensing2} (e.g. $0.02-0.08 \times 10^{-18}\,\mr{N m}$ for ATP synthase (F1-ATPase) \cite{torqueSensing3} or $\sim 0.01 \times 10^{-18}\,\mr{N m}$ for  RNA polymerase \cite{torqueSensing2}). The rotational trap stiffnesses we have measured are $0-1.2 \times 10^{-18}\,\mr{N m/rad}$ and can be controlled by varying particle size and electric field magnitude and frequency. These stiffness values are within the range of those in magnetic tweezer experiments with superparamagnetic micro-beads tethered to a surface \cite{torqueSensing1}. The torque resolution for sensing depends on the uncertainty of the angle detection via rotational tracking. We achieve $\sim 0.2\,\mr{degrees}$ ($\sim 0.004\,\mr{rad}$) in our experiments (see Appendix \ref{App:ImageProcessing}), meaning that our reduced rotational trap stiffnesses would allow sufficient resolution for biological torque sensing. Most current single-molecule force and torque spectroscopy techniques suffer from the disadvantage of coupled torque and force sensing \cite{torqueSensing2}. The use of our HOPG micro-particles together with appropriately applied magnetic and electric fields to independently control rotational trapping and translational confinement, could result in the future in extremely useful, decoupled force and torque sensing schemes.


\section*{Appendices}

\appendix

\sectionfont{\sffamily\large}

\section{Preparation of lipid-coated, dispersed HOPG micro-particles} \label{App:LipidCoating}

We have developed the following protocol to coat and solubilize micron-sized HOPG particles with lipid bilayers. Briefly, 1-palmitoyl-2-oleoyl-sn-glycero-3-phosphocholine (POPC, Avanti Polar Lipids) is dissolved in chloroform ($\mr{CHCl}_3$) and the resulting solution is dried under a $\mr{N}_2$ gas stream. The solvent-free lipid film is hydrated at a temperature ($\sim 60^{\circ} \mr{C}$) well above the lipid main phase transition ($-2^{\circ} \mr{C}$) with $20\,\mr{mM}$ aqueous solution of NaCl, resulting in a final lipid concentration of 0.6 mg/ml. To obtain small unilamellar lipid vesicles, this dispersion is briefly sonicated with a probe sonicator. Then the sample is centrifugated (30 minutes at 8000 rpm) to pellet down titanium probe particles and residual multilamellar vesicles. Small pieces ($\sim 1\,\mr{mm}^3$) of a HOPG crystal (ZYH grade, $3.5^{\circ}$ mosaic spread, from MikroMasch) are introduced in the NaCl aqueous lipid solution and sonicated in a small bath at $60^{\circ} \mr{C}$  for 5-30 minutes to induce the coating of the particles. Longer sonication times lead to a larger population of solubilised HOPG particles. The result is a clearly dispersed solution of HOPG particles. For our experiments, micron-sized flakes with average dimensions $\sim 2\,\mu\mr{m} \times 4\,\mu\mr{m} \times 7\,\mu\mr{m}$ are selected. 

\begin{figure}[h]
\centering
  \includegraphics[width=0.95\columnwidth]{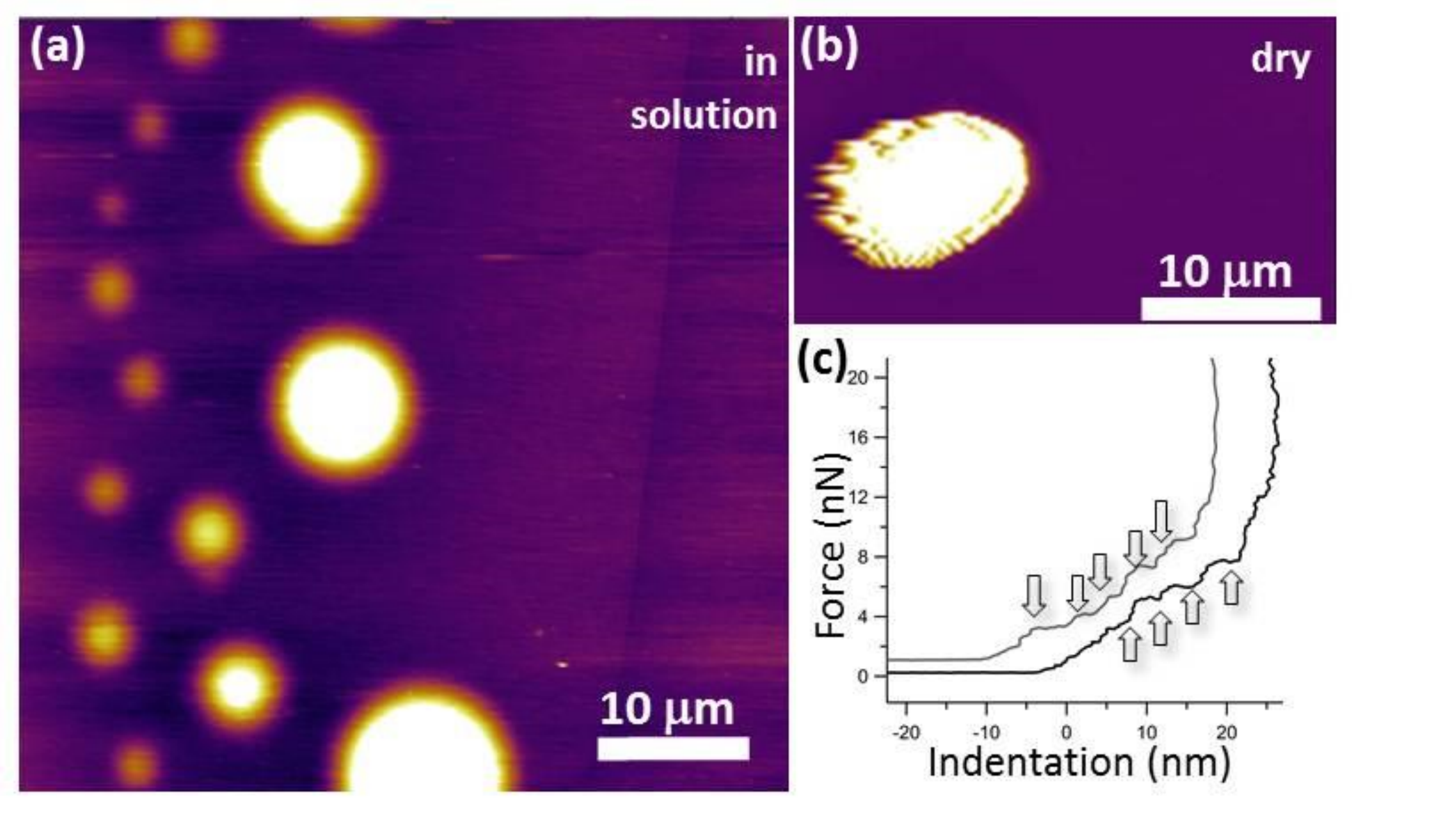}
  \caption{(a) Amplitude modulation (AM) AFM image of dispersed graphite flakes of different sizes coated with lipids in solution. They take a characteristic round shape. The height range is  20 nm. (b) AM-AFM image of a lipid-coated graphite flake imaged in air. (c) Force vs. indentation curves on two large flakes of figure (a), the arrows indicate the points where the AFM tip seems to penetrate a lipid bilayer, suggesting that the flakes are coated with several lipid bilayers.}
  \label{fig:AFM}
\end{figure}

Figure \ref{fig:AFM} shows atomic force microscopy (AFM) images of lipid-coated HOPG micro-particles of different sizes. The AFM image in solution [Fig. \ref{fig:AFM}(a)] shows very round particles, corresponding to the external lipid bilayers forming a vesicle-like structure. Force versus indentation curves [Fig. \ref{fig:AFM}(c)] reveal 4-5 indentation steps, suggesting that the particles are coated with several lipid bilayers.

\section{Image processing} \label{App:ImageProcessing}

Particles are first automatically detected on each frame in the sequence by means of a thresholding operation. This results in a connected region representing the particle mask, surrounded by a background region. The ellipse that best fits the particle-mask shape is then found [particles are approximated as ellipsoids, see last frame in Fig. \ref{fig:magneticOrientationImages}(C)] and the orientation angle $\varphi$ of the particle is obtained. The uncertainty of our angle measurements is $\sim 0.2\,\mr{degrees}$. This is determined using samples of fixed HOPG micro-flakes and calculating the standard deviation of the distribution of angles extracted with our image processing algorithms over 20000 repetitions.

Particle-dimension estimates are obtained for each micro-flake from the lengths of the minor and major axes of the fitted ellipses. For the 10 different individual lipid-coated HOPG micro-particles we have used in our measurements, the average dimensions are $\sim 2\,\mu\mr{m} \times 4\,\mu\mr{m} \times 7\,\mu\mr{m}$ (ellipsoid axis lengths with standard deviations $\pm \, 0.5\,\mu\mr{m}$).

\section{Correction for the variation with frequency of the actual voltage amplitude at the sample} \label{App:CorrectionFactor}

It is important to take into consideration impedance-mismatch effects in our electrical connections at the high frequencies (1-$70\,\mr{MHz}$) of the electric fields employed in our experiments. Our circuit components (radio frequency (RF) signal generator, amplifier, switch, coaxial cables) are all specified for an impedance $Z_0=50\,\Omega$. However, the impedance we measure for our electrode wires is $Z_\mr{w}\sim 100\,\Omega$, consistent with that of a parallel-wire transmission line made out of enamel-coated wires ($50\,\mu\mr{m}$ wire diameter, $150\,\mu\mr{m}$ wire distance). It is actually not possible to match this to $Z_0=50\,\Omega$ due to geometrical constraints. We minimise abrupt changes in impedance by carefully tapering connecting wire distances and use twisted wires where possible to avoid RF-noise pick up. The unavoidable impedance mismatch results in some reflections at connection junctions and cavity build-up effects which vary with the frequency of the signal.
\begin{figure}[h]
\centering
  \includegraphics[width=0.9\columnwidth]{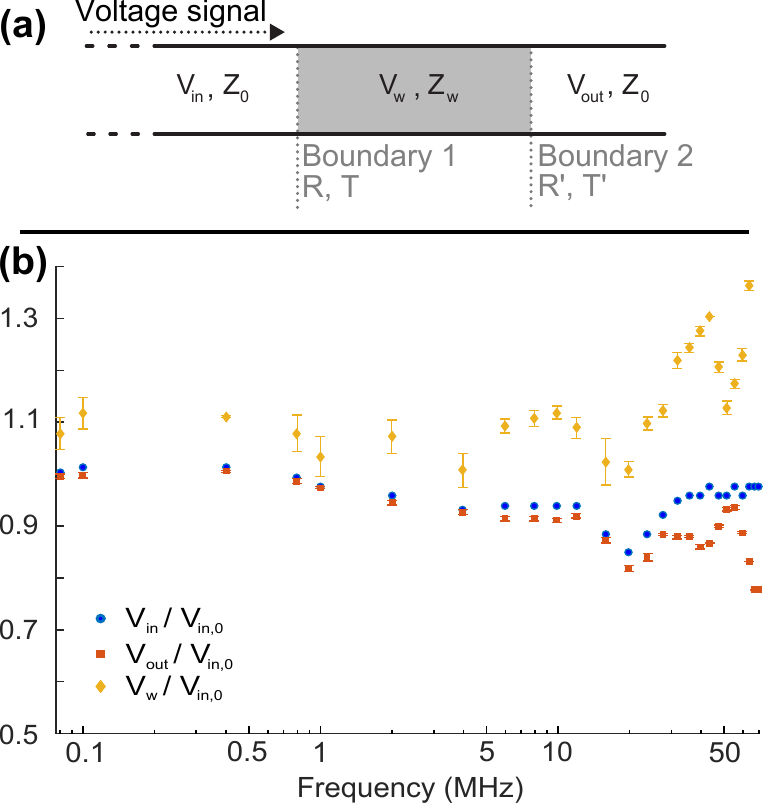}
  \caption{(a) Schematic of transmission line representing the thin electrode wires at the sample and connections to either side. (b) Measured values versus frequency of the input voltage amplitude into the wires, $V_\mr{in}$, voltage amplitude transmitted through them, $V_\mr{out}$, and calculated voltage amplitude at the sample, $V_\mr{w}$, all normalised to the input voltage at zero frequency, $V_\mr{in,0}$.}
  \label{fig:impedance}
\end{figure}

\begin{figure}[h]
\centering
  \includegraphics[width=0.8\columnwidth]{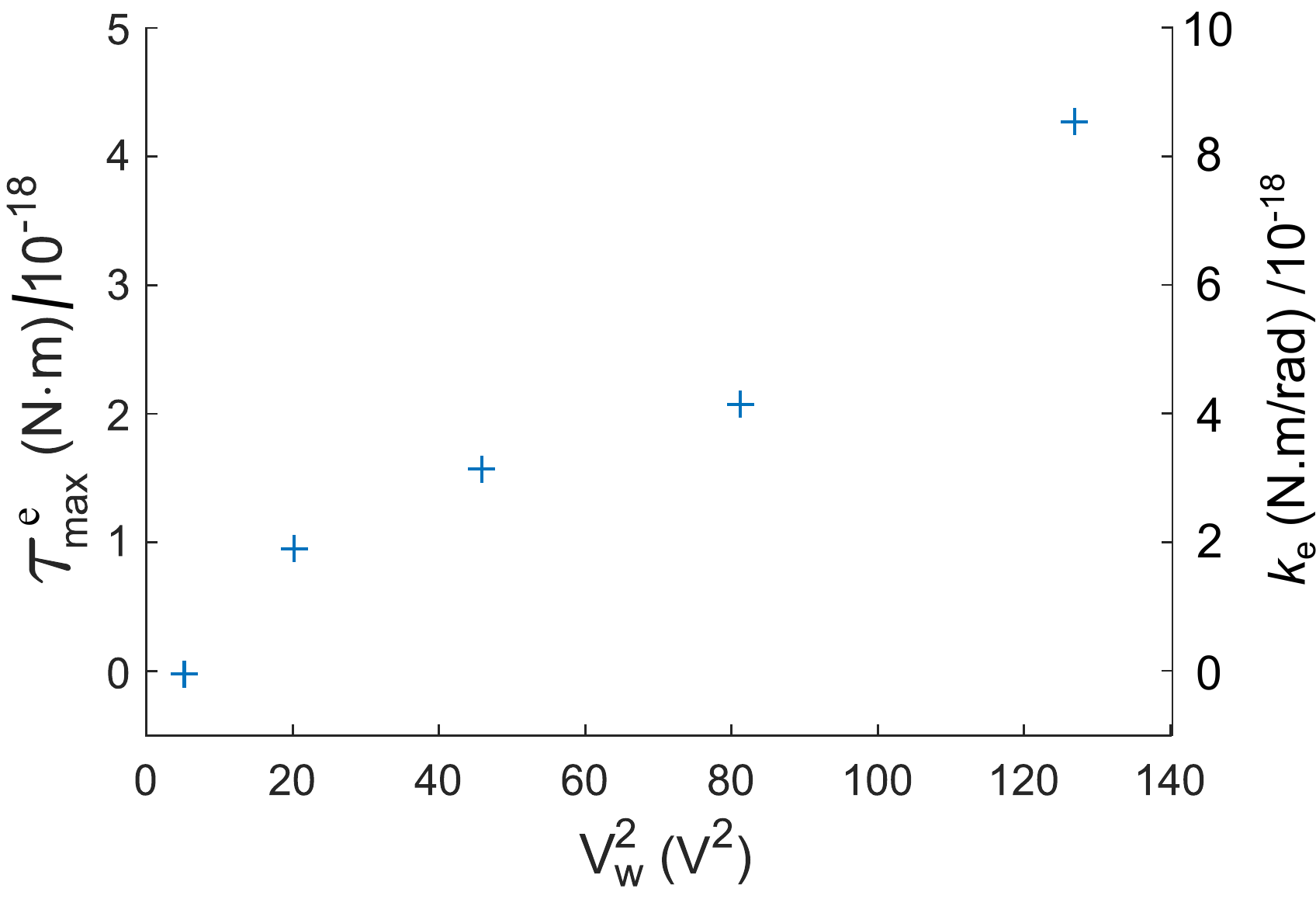}
  \caption{Measured maximum electric torque $\mathcal{T}_\mr{max}^\mr{e}$ and rotational trap stiffness $k_\mr{e}$ as a function of the squared voltage amplitude applied to the sample electrode wires, $V_\mr{w}^2$, for an electric field frequency of $20\,\mr{MHz}$, for an individual micro-flake.}
  \label{fig:torqueVsV}
\end{figure}

Additionally, our amplified input voltage signal presents some variation with frequency. We can measure the input alternating voltage amplitude, $V_\mr{in}$, into the electrode wires and the output voltage, $V_\mr{out}$, transmitted through them [see Fig. \ref{fig:impedance}(A)], as a function of frequency, and define $V_\mr{r}=V_\mr{out}/V_\mr{in}$. By considering voltage signal reflection and transmission through two subsequent boundaries with impedances $Z_0 \rightarrow Z_\mr{w}$ and $Z_\mr{w} \rightarrow Z_0$ at either side [see Fig. \ref{fig:impedance}(A)], we can calculate the voltage amplitude at the sample as $V_\mr{w}=V_\mr{in}[2Z_\mr{w}/(Z_0+Z_\mr{w})]$. The factor $[2Z_\mr{w}/(Z_0+Z_\mr{w})]$ corresponds to the voltage transmission coefficient through the input $Z_0 \rightarrow Z_\mr{w}$ boundary. $Z_\mr{w}$ can be calculated as $Z_\mr{w}=(Z_0/V_\mr{r})[(2-V_\mr{r})+2 \sqrt{1-V_\mr{r}}]$ using the measured value of $V_\mr{r}$ at each frequency. We find no appreciable difference between voltage measurements in solution and in dry samples, within our measurement uncertainties, owing to the fact that our thin sample wires are insulated with an enamel coating. Figure \ref{fig:impedance}(B) shows the measured $V_\mr{in}$ and $V_\mr{out}$ amplitudes and calculated $V_\mr{w}$ as a function of frequency, normalised to the value of the input voltage amplitude at zero frequency, $V_\mr{in,0}$ (typically $\sim 2.3\,\mr{V}$). Five measurements are taken at each frequency, re-soldering the thin electrode wires each time, to check reproducibility upon re-connection. Mean values are shown as data points, with standard errors as error bars. Since, as given by Eqn. (\ref{eqn:Tmax}) with $E_0 = V_\mr{w}/d$ and as shown by the data in Fig. \ref{fig:torqueVsV}, the electric torque is proportional to the square of the electric field modulus and, therefore, proportional to $V_\mr{w}^2$, a correction factor $(V_\mr{w}/V_\mr{in,0})^{-2}$ is used in our data analysis to multiply and scale our measured values of electric torque as a function of frequency, in order to correct for the above-mentioned effects. 

\section{Formulas for oblate ellipsoids} \label{App:OblateEllipsoidFormulas}

Approximating our HOPG micro-flakes as oblate ellipsoids with semi-axes $a < b \approx c$, with $a$ perpendicular to the graphene planes and $b$ and $c$ parallel to them, we can use analytical expressions to calculate the rotational friction coefficient, $C_{\parallel}$, for rotations of the ellipsoid around any of its in-plane axes \cite{PerrinBrownianEllipsoid}:
\begin{equation}\label{eqn:RotFrictionCoeff}
	C_{\parallel} = \frac{32 \pi \eta}{3}  \frac{\left(a^4-b^4\right)}{\left(2a^2-b^2\right)S_\mr{obl}-2a}  \, ,
\end{equation}
where $\eta$ is the dynamic viscosity of the solution and the geometrical factor $S_\mr{obl}$ for oblate ellipsoids takes the form:
\begin{equation}\label{eqn:Soblate}
	S_\mr{obl} = \frac{2}{\sqrt{b^2-a^2}} \arctan\left(\frac{\sqrt{b^2-a^2}}{a}\right)  \, .
\end{equation}

\sectionfont{\sffamily\Large}

\section*{Acknowledgments}
The authors thank Jonathan G. Underwood, Peter Barker and Phil H. Jones for very useful discussions. The authors would like to acknowledge EPSRC funding for this work.






\bibliography{HOPGorient} 
\bibliographystyle{rsc} 

\end{document}